\documentclass[%
 reprint,
superscriptaddress,
 amsmath,amssymb,
 aps,
]{revtex4-2}

\usepackage{bm}%
\usepackage{float}
\usepackage{graphicx}% Include figure files
\usepackage{dcolumn}% Align table columns on decimal point
\usepackage{bm}%
\usepackage{hyperref}
\usepackage{amsmath}
\usepackage{amsfonts}
\usepackage{mathtools}
\usepackage{fixme}
\usepackage{color}
\usepackage{wasysym}

\begin{document}

%\preprint{APS/123-QED}

\title{Critical phenomena in light-matter systems with collective matter interactions}

\author{Ricardo Herrera Romero}
\affiliation{Departamento de F\'isica, Universidad Aut\'onoma Metropolitana-Iztapalapa, Av. Ferrocarril San Rafael Atlixco 186, C.P. 09310, CDMX, Mexico.}
\author{Miguel Angel Bastarrachea-Magnani}%
\email{bastarrachea@xanum.uam.mx}
\affiliation{Departamento de F\'isica, Universidad Aut\'onoma Metropolitana-Iztapalapa, Av. Ferrocarril San Rafael Atlixco 186, C.P. 09310, CDMX, Mexico.}
\author{Román Linares}
\affiliation{Departamento de F\'isica, Universidad Aut\'onoma Metropolitana-Iztapalapa, Av. Ferrocarril San Rafael Atlixco 186, C.P. 09310, CDMX, Mexico.}

\begin{abstract}
We study the quantum phase diagram and the onset of quantum critical phenomena in a generalized Dicke model that includes collective qubit-qubit interactions. By employing semiclassical techniques, we analyze the corresponding classical energy surfaces, fixed points, and the smooth Density of States as a function of the Hamiltonian parameters to determine quantum phase transitions in either the ground (QPT) or excited states (ESQPT). We unveil a rich phase diagram, the presence of new phases, and new transitions that result from varying the strength of the qubits interactions in independent canonical directions. We also find a correspondence between the phases emerging due to qubit interactions and those in their absence but with varying the strength of the non-resonant terms in the light-matter coupling. We expect our work to pave the way and stimulate the exploration of quantum criticality in systems combining matter-matter and light-matter interactions.
\end{abstract}

\maketitle

%%%%%%%%%%%%%%%%%%%%%%%%%%%%%%%%%%%%%%%%%%

%%%%%%%%%%%%%%%%%%%%%%%%%%%%%%%%%%%%%%%%%%

\section{Introduction}
%%%%%%%%%%%%%%%%%%%%%%%%%%%%%%%%%%%%%%%%%%
%%%%%%%%%%%%%%%%%%%%%%%%%%%%%%%%%%%%%%%%%%

%Motivation for the study of QPT.
Quantum phase transitions (QPT) are generally defined as the sudden change in the properties of the ground-state of a quantum system as a function of a control parameter. They possess an essential role in modern physics, especially in studying many-body quantum systems, quantum information, and quantum control~\cite{Sachdev99,Carr10}. The active interest in quantum critical phenomena during the last two decades stems from their impact on the spectral features and dynamics of complex quantum systems, leading, e.g., to the development of new concepts such as that of Excited-State Quantum Phase Transition (ESQPT), that is meant to explain the consequences of the propagation of critical behavior from the ground-state to the rest of the spectrum of a quantum system~\cite{Caprio08,Stransky14,Stransky15,Cejnar2021}; and that of Dynamical Quantum Phase Transition (DQPT), seeking to fathom the onset of criticality exhibited in non-equilibrium phenomena~\cite{Klinder15,Heyl2018,Link2020}. Typically, understanding Quantum Phase Transitions depends on the specific system of study. Thus, the field remains an open challenge with exciting avenues, striving to reach a general framework to describe the interplay between many-body properties, strong interactions, and critical phenomena. 

% Dicke Hamiltonian, applications, and classical limit. 
A paradigmatic example of a QPT is the Dicke Hamiltonian's superradiant phase transition~\cite{Hepp73,Wang73}. The Dicke model describes a collection of atoms within the two-level approximation interacting with a single-mode radiation field inside a cavity~\cite{Dicke54}. The superradiant QPT is characterized by a non-zero expectation value of the photon number when the light-matter strength reaches a critical value in the thermodynamic limit. Because it describes the collective degrees of freedom of a set of two-level systems (qubits), the Dicke Hamiltonian offers a general description of the spin-boson interaction. Also, it constitutes a paradigmatic example for the study of the ultra-strong coupling (USC) regime~\cite{FornDiaz2019,FriskKockum2019,MarquezPeraca2020} Consequently, the model has found a great reception in the description of several setups, mainly in the context of quantum information~\cite{Garraway2011,Kirton2019,LeBoite2020,Larson2021}. During the last years, it has been experimentally realized in a broad range of tunable systems, from Bose-Einstein condensates in optical lattices~\cite{Schneble03,Baumann10,Baumann11,Klinder15,Keeling2010}, superconducting qubits~\cite{Blais04,Casanova10,Mezzacapo14} to cavity-assisted Raman transitions~\cite{Dimer2007,Baden14}. Not only formal derivations of Dicke-like Hamiltonians have been found in the framework of ultracold atoms in optical lattices~\cite{Nagy10,Liu11,Yuan17}, and for superconducting qubits~\cite{Jaako2016,Yang17,DeBernardis2018,Pilar2020}, but superradiant effects have also been proposed in nuclear physics~\cite{Auerbach11}, solid-state physics~\cite{Cong16}, bidimensional materials~\cite{Hagenmuller12,Chirolli12}, and quantum dots~\cite{Scheibner07}, among others. Moreover, its algebraic simplicity allows one to employ it as a test bed for studying critical features in the spectrum, several topics relevant to quantum information like quantum chaos and quantum correlations~\cite{Lambert04,Brandes05,Vidal06}, and the quantum-classical correspondence~\cite{Deaguiar1991,Deaguiar1992,Furuya1992,Bastarrachea2015}. The last is possible because the Hamiltonian can be mapped to only two relevant degrees of freedom (those of the boson and the collective spin), so there is a well-defined classical limit. This feature has raised questions about the nature of the superradiant QPT, being deemed as a mean-field QPT due to its classicality and smallness of quantum fluctuations, which are the ones driving phase transitions at low temperatures
~\cite{Larson17}. 

% Qubit-qubit interactions.
An additional feature in those systems where the Dicke model finds application is the possibility to build up collective qubit-qubit interactions. Previous works have addressed this problem by considering, e.g., dipolar interactions in atomic systems~\cite{Chen2006}, shifts due to the Stark effect in optomechanical setups~\cite{Abdel-Rady2017,Salah2018}, Josephson dynamics in a two-component BEC~\cite{Chen2007,Rodriguez2011,Sinha2019}, or the onset of chaos~\cite{Rodriguez2018,Sinha2020}. Two general results stand due to the presence of matter interactions: the prediction of a first-order phase transition~\cite{Lee2004,Chen2008,Chen2010,Rodriguez2011,Rodriguez2018}, the shift of the critical coupling of the standard superradiant phase transition~\cite{Chen2006,Jaako2016} —including its possible suppression—, and a richer phase diagram~\cite{Robles2015,Rodriguez2018}. Despite previous studies of the ground-state properties in this system, the interplay of these collective interactions on the spectral properties of the Hamiltonian, as well as the understanding of the different energy domains marked by the presence of quantum phase transitions, has not been done exhaustively.   

% Our contribution
In this work, we are interested in studying the critical behavior of a generalized Dicke Hamiltonian that includes collective qubit-qubit interactions. It will constitute an example of the intriguing combination between matter-matter interactions and (ultra) strong light-matter ones and the rich phase diagrams they can produce. Unlike previous works, here we will add a general combination of non-linear interactions in the form of quadratic terms in the collective pseudo-spin operators to the standard Dicke Hamiltonian in all the $x$-, $y$- and $z$-directions. Then, we perform a standard semiclassical analysis to obtain the behavior of energy surfaces, the ground-state energy, and the Density of States (DoS) as a function of the Hamiltonian parameters. This exploration allows us, from a unified perspective, to obtain indicators of critical quantum behavior, i.e., both QPT and ESPQT, as has been done previously in other works~\cite{Bastarrachea2014a,Bastarrachea2016,Rodriguez2018}. We offer a general overview that unifies some results of previous works and found new behavior unlocked by the unique interplay between the different directions of the interactions.

% Organization
The article is organized as follows. In Sec.~\ref{sec:2} we present the generalized Dicke Hamiltonian, including qubit-qubit interactions. Next, in Sec.~\ref{sec:3} we discuss the corresponding classical Hamiltonian obtained via coherent states, the Hamilton equations of motion, and the fixed points, commonly associated with critical behavior in the ground-state of the related quantum system. Also, we present an overview of the classical energy surfaces and classify the different phases according to the ground-state properties. In Sec.~\ref{sec:4} we calculate the semi-classical density of states (DoS) to identify the ESQPT and the spectral domains. Finally, in Sec.~\ref{sec:5} we present our conclusions. We include several appendices with details on the calculations.  

%%%%%%%%%%%%%%%%%%%%%%%%%%%%%%%%%%%%%%%%%%%%%
%%%%%%%%%%%%%%%%%%%%%%%%%%%%%%%%%%%%%%%%%%%%%
\section{Generalized Dicke Hamiltonian}
\label{sec:2}
%%%%%%%%%%%%%%%%%%%%%%%%%%%%%%%%%%%%%%%%%%%%%
%%%%%%%%%%%%%%%%%%%%%%%%%%%%%%%%%%%%%%%%%%%%%

We study a generalized Dicke Hamiltonian that includes collective qubit-qubit interactions 
\begin{gather} \label{eq:hd}
        \hat{H}_{D}=\hat{H}_{0}+\hat{H}_{\text{I}}+\hat{H}_{\text{qq}},
\end{gather}
where
\begin{gather} 
        \hat{H}_{0}=\omega 
        \hat{a}^{\dagger}\hat{a}+\omega_{0}\hat{J}_{z},
        \nonumber
        \\ \nonumber
                \hat{H}_{\text{I}}=\frac{\gamma}{\sqrt{N}}\left[(\hat{a}\hat{J}_{+}+\hat{a}^{\dagger}\hat{J}_{-})+\xi(\hat{a}^{\dagger}\hat{J}_{+}+\hat{a}\hat{J}_{-})\right], \\ \nonumber
        \hat{H}_{\text{qq}}=\frac{1}{N}\left(\eta_{x}\hat{J}_{x}^{2}+\eta_{y}\hat{J}_{y}^{2}+\eta_{z}\hat{J}^{2}_{z}\right).
\end{gather}
The first term denotes the non-interacting Hamiltonian $\hat{H}_0$, the second one is the $\hat{H}_{\text{I}}$ usual spin-boson interaction, and the last one contains the qubit-qubit interactions $\hat{H}_{\text{qq}}$. Here, $\hat{a}^{\dagger}$ ($\hat{a}$) is the creation (annihilation) boson operator, and $\hat{J}_{z,x,y}$ are the pseudo-spin operators representing the collective degrees of freedom of the set of $N$ qubits, which follow the rules of the $\text{su(2)}$-algebra. Generally, a set of $N$ qubits is spanned into a $2^{N}$ dimensional Hilbert space; however, as we are interested in describing the collective degrees of freedom, it suffices to work in the totally symmetric subspace corresponding to $j=N/2$, where $j(j+1)$ is the eigenvalue of the pseudo-spin length operator $\hat{\mathbf{J}}^{2}$. Thus, the dimension of the Hilbert space is reduced to only $N+1$, where the collective ground state lies. The Hamiltonian parameter set is given by $\omega$, $\omega_0$, and $\gamma$ are the boson frequency, the qubit energy splitting, and the spin-boson interaction. Additionally, we have $\eta_{i}$ with $i=x,y,z$, the collective qubit-qubit couplings in each direction. Depending on the setup, one can grant a specific meaning to the interactions in the $z$ and its perpendicular directions. An intuitive approach comes from interacting Bose-Einstein condensates in a two-site trap and Josephson effects. There, $\hat{J}_{z}$ is related to a relative population of particles in the condensates and $\hat{J}_{x}$ ($\hat{J}_{y}$) to ladder operators and relative phases between them. Thus, $\eta_{z}$ and $\eta_{x}$ ($\eta_{y}$) represent the strength of collective on-site and between neighboring sites interactions (hopping effects), respectively~\cite{Chen2007}. Otherwise, interactions from the $x$ and $y$ directions arise from dipolar coupling in atomic setups~\cite{Joshi1991,Chen2006} or interactions between superconducting qubits~\cite{Jaako2016,DeBernardis2018,Pilar2020}.

The Hamiltonian in Eq.~\ref{eq:hd} possesses several well-known limits. In the absence of qubit-qubit interactions ($\eta_{i}=0$ for $i=x,y,z$), one recovers the standard light-matter interaction. The parameter $\xi$ takes the system from the integrable Tavis-Cummings model ($\xi=0$)~\cite{TC1968}, which describes a system in the strong coupling regime under the Rotating-Wave Approximation (RWA), to the standard, non-integrable Dicke model ($\xi=1$) typically describing the USC~\cite{Dicke54}. In both limits, the superradiant QPT takes place when the light-matter coupling crosses the critical value $\gamma_{\xi+}=\sqrt{\omega\omega_{0}}/(1+\xi)$. For values below the coupling ($\gamma<\gamma_{\xi+}$) the system is in a normal phase, characterized by a zero-average of photon population in the thermodynamical limit ($\bar{n}=\langle \hat{a}^{\dagger}\hat{a}\rangle/N=0$), while for $\gamma_{\xi+}>\gamma$ one finds a finite photon number $\bar{n}\neq 0$, thus called superradiant phase. Besides, the Hamiltonian exhibits two ESQPTs~\cite{PerezFernandez2011,Brandes13,Bastarrachea2014a,Bastarrachea2014b}, which are identified as non-analyticities in the derivative of the smooth DoS as a function of energy in the thermodynamic limit. One at a critical energy $E_{-}^{(c1)}/\omega_{0}j=\epsilon_{-}^{(c1)}=-1$ that only appears in the superradiant phase (characterized by a logarithmic divergence in the derivative of the DoS as energy increases), and a second one at $\epsilon_{+}^{(c2)}=+1$ that appears for every coupling as a jump singularity (a step function in the DoS derivative)~\cite{Stransky2016} and is related to the saturation of the collective qubit Hilbert space. 

A finite value of $\xi\in(0,1)$ leads to the generalized or extended Dicke model instead~\cite{Deaguiar1992,Bastarrachea2014a,Kloc2017,Cejnar2021}. There, a new superradiant phase appears whose critical point occurs at $\gamma_{\xi-}=\sqrt{\omega\omega_{0}}/(1-\xi)$~\cite{Bastarrachea2016,Kloc2017,Stransky2017b}. While in the TC model $\gamma_{0+}=\gamma_{0-}$ and the new phase is equal to the standard one; in the Dicke model $\gamma_{1-}\rightarrow \infty$, so it becomes not observable. The ESPQTs predicted in the Dicke model persist in the generalized one; the only difference is that the ESQPT changes its type from a logarithmic singularity in the derivative of the smooth DoS to a step function with a downward jump from lower to higher energies in the interval $\gamma_{\xi+}<\gamma<\gamma_{\xi-}$~\cite{Stransky14,Stransky2017b,Cejnar2021}. On the other hand, in the absence of light-matter interaction, the boson is decoupled from the collective spin. Then, one gets a version of Lipkin-Meshkov-Glick Hamiltonian (LMG)~\cite{Lipkin65,Meshkov65,Glick65}, a well-known model with one degree of freedom originally coming from nuclear physics that nowadays is connected to Josephson junctions and cold atoms in optical lattices~\cite{Chen2009}. Critical phenomena in the LMG model have been extensively studied, and the system exhibits both a first-order and a second-order QPT, as well as ESQPTs~\cite{Dusuel04,Dusuel05,Castanos06,Heiss05,Leyvraz05,Heiss06,Ribeiro08,Engelhardt15,GarciaRamos17}, to cite some works. Naturally, we expect that the generalized Dicke Hamiltonian inherits critical features from the LMG model by including the qubit-qubit interactions.    

%%%%%%%%%%%%%%%%%%%%%%%%%%%%%%%%%%%%%%%%%%
%%%%%%%%%%%%%%%%%%%%%%%%%%%%%%%%%%%%%%%%%%
\section{Classical corresponding Hamiltonian}
\label{sec:3}
%%%%%%%%%%%%%%%%%%%%%%%%%%%%%%%%%%%%%%%%%%
%%%%%%%%%%%%%%%%%%%%%%%%%%%%%%%%%%%%%%%%%%

A classical Hamiltonian can be obtained by taking the expectation value of Eq.~\ref{eq:hd} over a tensor product of Glauber $|z\rangle$ and Bloch $|w\rangle$ coherent states as trial states~\cite{Deaguiar1992,Bastarrachea2014a,Bastarrachea2014b,Bastarrachea2015,Chavez2016}, 
where $|0\rangle$ and  $|j,-j\rangle$ are the boson and pseudo-spin vacuum states, respectively~\cite{Gilmore1990}, 
\begin{gather}
|z\rangle\otimes|w\rangle=\frac{e^{-|z|^2/2}}{(1+|w|^2)^{j}}e^{z\hat{a}^{\dagger}}e^{w\hat{J}_{+}}|0\rangle\otimes|j,-j\rangle.
\end{gather}
By dividing over $j$ we obtain
\begin{gather}
H_\text{cl}^{(\xi)}(z,w)  =j^{-1}\langle z,w|\hat{H}_{D}|z,w\rangle\\ \nonumber=
\omega|z|^{2}-\left(\frac{1-|w|^{2}}{1+|w|^{2}}\right)\left[\omega_{0}-\frac{\eta_{z}}{2}\left(\frac{1-|w|^{2}}{1+|w|^{2}}\right)\right]\\ \nonumber+
\frac{1}{2\left(1+|w|^{2}\right)^{2}}\left[\left(\eta_{x}-\eta_{y}\right)\left(w^{2}+\bar{w}^{2}\right)+\left(\eta_{x}+\eta_{y}\right)2w \bar{w}\right]\\ \nonumber+ \frac{\gamma\left(z+\bar{z}\right)\left(w+\bar{w}\right)}{\sqrt{2}(1+|w|^{2})}.
\end{gather}
Instead of employing the complex numbers $z$ and $w$, it is more convenient to use canonical classical variables $(q,p)$ and $(j_z,\phi)$ for the boson and spin spaces, respectively. Here $z=\sqrt{j/2}\left(q+ip\right)$ and $w=\sqrt{(1+j_z)/(1-j_z)}e^{-i\phi}$. Additionally, due to the fixed value of the pseudospin lenght $j=N/2$ we have $j_{x}=\sqrt{1-j_{z}^{2}}\cos\phi$ and $j_{y}=\sqrt{1-j_{z}^{2}}\sin\phi$. In this manner, we obtain a classical generalized Dicke Hamiltonian that reads
\begin{gather}
    H_{cl}^{(\xi)}=\frac{\omega}{2}(q^{2}+p^{2})+j_{z}\left(\omega_{0}+\frac{\eta_{z}j_{z}}{2}\right)\\ \nonumber+
    \frac{1}{2}\left(1-j^{2}_{z}\right)\left(\eta_{x}\cos^{2}\phi+\eta_{y}\sin^{2}\phi\right)\\ \nonumber
    +\gamma\sqrt{1-j^{2}_{z}}\left[(1+\xi)q\cos\phi-(1-\xi)p\sin\phi\right].
\end{gather}

To characterize the energy surfaces and identify the critical behavior, we will need the equations of movement. The Hamilton equations are
\begin{gather}\label{eq:he1}
    \dot{q}=\frac{\partial H_{cl}^{(\xi)}}{\partial p}=\omega p-\gamma\sqrt{1-j^{2}_{z}}(1-\xi)\sin\phi.
\end{gather}
\begin{gather} \label{eq:he2}
    \dot{p}=-\frac{\partial H_{cl}^{(\xi)}}{\partial q}=-\omega q-\gamma\sqrt{1-j^{2}_{z}}(1+\xi)\cos\phi,
\end{gather}
\begin{gather}\label{eq:he3}
   \dot{\phi}=\frac{\partial H_{cl}^{(\xi)}}{\partial j_{z}}=\omega_{0}+\eta_{z}j_{z}-j_{z}(\eta_{x}\cos^{2}\phi+\eta_{y}\sin^{2}\phi)
   \\ \nonumber
    -\frac{\gamma j_{z}}{\sqrt{1-j^{2}_{z}}}\left[(1+\xi)q\cos\phi-(1-\xi)p\sin\phi\right],
\end{gather}
\begin{gather}\label{eq:he4}
    \dot{j_{z}}=-\frac{\partial H_{cl}^{(\xi)}}{\partial \phi}=\left(1-j^{2}_{z}\right)(\eta_{x}-\eta_{y})\cos\phi\sin\phi+
    \\ \nonumber
    \gamma\sqrt{1-j^{2}_{z}}\left[(1+\xi)q\sin\phi+(1-\xi)p\cos\phi\right].
\end{gather}
In App.~\ref{app:a} we present the Hamilton equations for $\xi=0$ and $\xi=1$. 

%%%%%%%%%%%%%%%%%%%%%%%%%%%%%%%%%%%%%%%%%%
%%%%%%%%%%%%%%%%%%%%%%%%%%%%%%%%%%%%%%%%%%
\section{Energy surfaces and their extrema}
%%%%%%%%%%%%%%%%%%%%%%%%%%%%%%%%%%%%%%%%%%
%%%%%%%%%%%%%%%%%%%%%%%%%%%%%%%%%%%%%%%%%%

In this section we obtain the fixed, stationary or equilibrium points $(q_s,p_s,j_{zs},\phi_s)$ of the energy surface $H_{cl}(q_s,p_s,j_{zs},\phi_s)$ from Hamilton equations. They ease characterizing the system's different quantum phases and transitions as a function of the Hamiltonian parameters, employing that the minimum of the energy surface can be identified with the ground-state energy~\cite{Castanos2009}. In this case, the thermodynamic limit coincides with the classical limit because we have an effective Planck’s constant given by $\hbar_{eff}=\hbar/N$. So, as $N\rightarrow\infty$, $\hbar_{eff}\rightarrow 0$~\cite{Ribeiro2006}. Thus, to find the fixed points, we make Hamilton Eqs.~\ref{eq:he1} to~\ref{eq:he4} equal to zero. From the first two, we obtain a pair of equations defining the quadratures
\begin{gather} \label{eq:pq}
    p_s=\frac{\gamma}{\omega}\sqrt{1-j_{zs}^2}(1-\xi)\sin\phi_s,\,\,\,\mbox{and}\\ \nonumber
    q_s=-\frac{\gamma}{\omega}\sqrt{1-j_{zs}^2}(1+\xi)\cos\phi_s,
\end{gather}
Next, we can insert them into Eqs.~\ref{eq:he3} and~\ref{eq:he4} to get a second pair of equations that set the atomic (collective spin) variables
\begin{gather} \label{eq:c1}
\omega_0+j_{zs}\left\{\frac{}{}\eta_{z}-\left(\eta_{x}\cos^{2}\phi_{s}+\eta_{y}\sin^{2}\phi_{s}\right)\right.
\\ \nonumber
\left.+\frac{\gamma^{2}}{\omega}\left[(1+\xi)^{2}\cos^{2}\phi_{s}+(1-\xi)^{2}\sin^{2}\phi_{s}\right]\right\}=0,
\end{gather}
\begin{gather} \label{eq:c2}
(1-j_{zs}^{2})\cos\phi_{s}\sin\phi_{s}\mbox{x}\\\nonumber
\left\{(\eta_x-\eta_y)-\frac{\gamma^{2}}{\omega}\left[(1+\xi)^{2}-(1-\xi)^{2}\right]\right\}=0.
\end{gather}

We observe that Eqs.~\ref{eq:c1} and~\ref{eq:c2} are enough to determine the general conditions to find the fixed points. To better visualize the energy surfaces we study throughout this work, we use Eqs.~\ref{eq:pq} and a new set of atomic variables, as described in App.~\ref{app:b}. Then, the energy surface is restricted to the atomic space, simplifying the identification of fixed points. 

%%%%%%%%%%%%%%%%%%%%%%%%%%%%%%%%%%%%%%%%%%
%%%%%%%%%%%%%%%%%%%%%%%%%%%%%%%%%%%%%%%%%%
\subsection{Deformation of the normal phase}
%%%%%%%%%%%%%%%%%%%%%%%%%%%%%%%%%%%%%%%%%%
%%%%%%%%%%%%%%%%%%%%%%%%%%%%%%%%%%%%%%%%%%

Two fixed points that exist for every value of the Hamiltonian parameters. They come from Eq.~\ref{eq:c2}, when one makes $j_{zs}=\pm 1$. From Eq.~\ref{eq:pq}, at these values, one automatically gets that $p_{s}=q_{s}=0$, where $\phi_{s}$ is left indeterminate given that they coincide with the poles of the unitary Bloch sphere. The coordinates of these stationary points are 
\begin{gather}
     (p_{s},q_{s},j_{zs},\phi_{s})=(0,0,\pm 1,\text{indeterminate})
\end{gather}
and their energy is given by
\begin{gather}
    \epsilon_{\pm}=\pm 1+\frac{\eta_{z}}{2\omega_{0}}
\end{gather}
In the normal phase the stationary point at $j_{zs}=-1$ is a stable, absolute minimum~\cite{Bastarrachea2014a}. It corresponds to the lowest energy value of the system, marking the quantum ground-state energy. Because the expectation value of the photon number at the ground state $\langle \text{g.s.}|\hat{a}^{\dagger}\hat{a}|\text{g.s.}\rangle/N$ is in general proportional to $|z|^{2}=q^{2}+p^{2}$, it characterizes the quantum features of each phase. At the fixed point $j_{z}=-1$ we have that $q_{s}=p_{s}=0$, so we speak of a normal phase, distinguished by the absence of a strong-correlated light-matter quantum state that could lead, e.g., to a collective emission of photons. On the other hand, the point at $j_{zs}=+1$, which always belongs to a higher energy domain, is typically an unstable fixed point~\cite{Pilatowsky2020}. Because this point signals the maximum energy of the pseudospin (given that $|j_{zs}|\leq 1$), the entire phase space associated with the Bloch sphere becomes available for the pseudospin dynamics. Any additional energy will only increase the boson energy. Thus, it marks the onset of an ESQPT~\cite{Brandes13,Bastarrachea2014a}. 

The existence of a single global minimum in the standard normal phase of both the TC and the Dicke models is followed by energy surfaces that are invariant under $\phi$ rotations. This is connected to the conservation of the total number of excitations operator $\hat{\Lambda}=\hat{a}^{\dagger}\hat{a}+\hat{J}_{z}+j\hat{\mathbb{I}}$, as in the normal phase the system is virtually decoupled. In this case, the shape of the potential corresponds to a single well, as shown in Fig.~\ref{fig:1} (c3). However, even though the nature of the fixed points does not change, when $\eta_{x}-\eta_{y}\neq 0$, the energy surface becomes deformed thanks to the influence of interactions in $x$ and $y$ directions, and the rotational symmetry is broken at higher energies. We call this situation the {\it deformed normal phase}. The energy surfaces corresponding to this situation are shown, for example, in Fig.~\ref{fig:1} (c4) and (c7), where either the interactions in the $y$ or $x$ directions are present. Later, once we identify the parameter domains for the other phases, we will explain how the surface stretches depending on the qubit interaction directions. 

Finally, we notice that the energy of the points at $j_{z}=\pm 1$ is invariant concerning the qubit interactions in the $x$ and $y$ directions. Still, it is uniformly shifted by the $z$-interactions as was noted before in previous works~\cite{YiXiang2013,Rodriguez2011,Rodriguez2018}. The normal phase is thus identified by the presence of only these two fixed points. Additional stationary points will emerge, and the point at $j_{z}=-1$ will change its type of extrema according to the onset of the other phases. Next, we solve Eqs.~\ref{eq:pq},~\ref{eq:c1}, and~\ref{eq:c2} to find the those points and the phases for three different situations: 1) the Tavis-Cummings limit ($\xi=0$), 2) the Dicke limit ($\xi=1$), and 3) for an arbitrary value of $\xi$.

%%%%%%%%%%%%%%%%%%%%%%%%%%%%%%%%%%%%%%%%%%
%%%%%%%%%%%%%%%%%%%%%%%%%%%%%%%%%%%%%%%%%%
\subsection{Tavis-Cummings limit}
%%%%%%%%%%%%%%%%%%%%%%%%%%%%%%%%%%%%%%%%%%
%%%%%%%%%%%%%%%%%%%%%%%%%%%%%%%%%%%%%%%%%%

By setting $\xi=0$ we cancel the non-resonant terms in the Hamiltonian ($\hat{a}^{\dagger}\hat{J}_{+}$ and $\hat{a}\hat{J}_{-}$). Hence, we recover a Tavis-Cummings model modified by the qubit-qubit interactions. It corresponds to the situation where the Rotating-Wave Approximation (RWA) holds~\cite{Klimov2009}. The Hamiltonian becomes
\begin{gather}
    H_{cl}^{(0)}=\frac{\omega}{2}(q^{2}+p^{2})+j_{z}\left(\omega_{0}+\frac{\eta_{z}j_{z}}{2}\right)+\\\nonumber\frac{1}{2}\left(1-j^{2}_{z}\right)\left(\eta_{x}\cos^{2}\phi+\eta_{y}\sin^{2}\phi\right)+ \\ \nonumber
    \gamma\sqrt{1-j^{2}_{z}}\left(q\cos\phi-p\sin\phi\right),
\end{gather}
and Eqs.~\ref{eq:c1} and~\ref{eq:c2} are in this case
\begin{gather} \label{eq:c1tc}
    \omega_{0}+j_{zs}\left(\eta_{z}-\left(\eta_{x}\cos^{2}\phi_{s}+\eta_{y}\sin^{2}\phi_{s}\right)+\frac{\gamma^{2}}{\omega}\right)=0, \\
    (1-j_{zs}^{2})(\eta_x-\eta_y)\cos\phi_{s}\sin\phi_{s}=0.
\end{gather}
We find five different solutions for the stationary points (including $j_{zs}=\pm1$). The other three conditions are given by the cases: $\cos\phi_{s}=0$ ($\sin\phi_{s}=\pm1$), $\sin\phi_{s}=0$ ($\cos\phi_{s}=\pm1$), and $\eta_{x}=\eta_{y}$.

%%%%%%%%%%%%%%%%%%%%%%%%%%%%%%%%%%%%%%%%%%
%%%%%%%%%%%%%%%%%%%%%%%%%%%%%%%%%%%%%%%%%%
\subsubsection{Superradiant-symmetric phase}
%%%%%%%%%%%%%%%%%%%%%%%%%%%%%%%%%%%%%%%%%%
%%%%%%%%%%%%%%%%%%%%%%%%%%%%%%%%%%%%%%%%%%
We start studying the situation when the interactions in the $x$ and $y$ directions are the same $\eta_{x}=\eta_{y}=\eta$. Eq.~\ref{eq:c1tc} becomes
\begin{gather}
    \omega_{0}+j_{zs}\left(\eta_{z}-\eta\right)+\frac{\gamma^{2}}{\omega}=0. 
\end{gather}
Thus, we can obtain $j_{zs}$ immediately as
\begin{gather}
    j_{zs}=-\frac{\omega_{0}}{\eta_{z}-\eta+\frac{\gamma^{2}}{\omega}}=-\frac{1}{f_{0}},\,\,\,
f_{0}=\frac{\Delta\eta_{zs}}{\omega_{0}}+f_{0+},
\end{gather}
where $f_{0+}=\gamma^{2}/\gamma_{0+}^{2}$, $\gamma_{0+}=\sqrt{\omega\omega_{0}}$ is the critical coupling of the superradiant phase in the standard TC model, and $\Delta\eta_{zs}=\eta_{z}-\eta$. Substituting the value of $j_{zs}$ in the definitions Eq.~\ref{eq:pq} we obtain the stationary points 
\begin{gather} 
     (p_{s},q_{s},j_{zs},\phi_{s})=\left(\frac{\gamma}{\omega}\sqrt{1-\frac{1}{f_{0}^{2}}}\sin\phi_{s},\right.\\\nonumber\left.-\frac{\gamma}{\omega}\sqrt{1-\frac{1}{f_{0}^{2}}}\cos\phi_{s},-\frac{1}{f_{0}},\text{indeterminated}\right).
\end{gather}
In other words, there is a continuous of fixed points, associated with the conservation of the total number of excitations which makes the standard TC Hamiltonian integrable. This leads to the standard result from the TC model where the energy surface takes the form of the Mexican hat potential~\cite{Bastarrachea2014a}, as shown in Fig.~\ref{fig:1} (c3). These points are valid when $f_{0}\geq 1$, so the value of $j_{zs}$ remains real. Thus, there is a critical coupling given by
\begin{gather}\label{eq:crittcz}
    \gamma_{0}^{c}=\gamma_{0+}\sqrt{1-\frac{\Delta\eta_{zs}}{\omega_{0}}},
\end{gather}
where we get the standard critical coupling of the TC model modified by a factor that depends on the qubit-qubit interactions (for $\Delta\eta_{zs}\geq 0$ the critical coupling $\gamma_{0}^{c}$ becomes zero). 
%Also, this phase is valid only if $\Delta\eta_{zs}\leq\omega_{0}$. 
This set of points has an energy
\begin{gather} \label{eq:es0}
    \epsilon_{s0\phi}=-\frac{1}{2}\left(f_{0}+\frac{1}{f_{0}}\right)+\frac{\eta_{z}}{2\omega_{0}}.
\end{gather}
This value is obtained from $\epsilon=H_{cl}(q_{s},p_{s},j_{zs},\phi_{s})/\omega_{0}$. 
As it can be straightforwardly seen, for this parameter domain $\epsilon_{s0\phi}<\epsilon_{-}$. So, if one calculates the Hessian matrix (see App.~\ref{app:c}), one can identify these points as a set of minima. Instead, for $f_{0}\geq 1$ the point at $j_{zs}=-1$ becomes a local maximum, while the one at $j_{zs}=+1$ remains the absolute maximum. We notice that the ground-state energy is continuous, so $\epsilon_{s\phi}=\epsilon_{-}$ at $f_{0}=1$. According to the form of ground-state energy, we have $q_{s}\neq 0$ and $p_{s}\neq 0$. As a result, the domain where $f_{0}\geq 1$ 
%and $\Delta\eta_{zs}\leq\omega_{0}$ 
is recognized as a superradiant phase. One of the major effects of the interactions with respect to the standard TC Hamiltonian is the shift in the critical coupling: both the interactions in the $z$ and $x$ ($y$) directions change it. Moreover, we notice that the critical coupling $\gamma_{0}^{c}$ becomes zero when $\eta_{z}-\eta=\omega_{0}$. This means that, for interacting values where $\eta_{z}-\eta\geq\omega_{0}$, there is only a superradiant phase, but not a normal phase for every value of the coupling! Thus, the onset of the superradiant phase can be suppressed or stimulated by choosing the right value of the relative interactions in the $z$ and perpendicular directions.

Interestingly, we observe that the rotational symmetry is not broken in this case because the qubit-qubit interactions are balanced in $x$ and $y$ directions ($\eta_{x}=\eta_{y}$). Hence, we call the quantum phase existing for $f_{0}\geq1$ and $\eta_{x}\neq\eta_{y}$
%, and $\Delta\eta_{zs}\geq\omega_{0}$% 
the superradiant-symmetric phase. Next, we consider the imbalanced case ($\eta_{x}\neq \eta_{y}$), which leads to two different, but symmetric to each other, superradiant phases.

%%%%%%%%%%%%%%%%%%%%%%%%%%%%%%%%%%%%%%%%%%
%%%%%%%%%%%%%%%%%%%%%%%%%%%%%%%%%%%%%%%%%%
\subsubsection{Superradiant-x phase}
%%%%%%%%%%%%%%%%%%%%%%%%%%%%%%%%%%%%%%%%%%
%%%%%%%%%%%%%%%%%%%%%%%%%%%%%%%%%%%%%%%%%%
Now, we consider $\eta_{x}\neq\eta_{y}$ ($\Delta\eta_{zx}\neq\Delta\eta_{zy}$) and $\cos\phi_s=\pm1$ ($\sin\phi_s=0$). Here, we get $p_{s}=0$, and from Eq.~\ref{eq:c1} we obtain
\begin{gather}
    j_{zs}=-\left(\frac{\eta_{z}-\eta_{x}}{\omega_{0}}+\frac{\gamma^{2}}{\omega\omega_{0}}\right)^{-1}=-\frac{1}{f_{0x}},\,\,\, \\ \nonumber
f_{0x}=\frac{\Delta\eta_{zx}}{\omega_{0}}+f_{0+},
\end{gather}
which looks exactly as in the previous case with a critical coupling given by
\begin{gather}\label{eq:critt0x}
    \gamma_{0x}^{c}=\gamma_{0+}\sqrt{1-\frac{\Delta\eta_{z,x}}{\omega_{0}}}.
\end{gather}
where $\Delta\eta_{zx}=\eta_{z}-\eta_{x}$. Similarly, for $\Delta\eta_{zx}\geq\omega_{0}$, the critical coupling $\gamma_{0x}^{c}$ becomes zero.
%Also, there is an extra condition given by $\Delta\eta_{zx}\leq\omega_{0}$. 
Unlike the previous case, however, there is not an infinite set of stationary points but only two degenerated ones. This is the most common case for all the superradiant phases we will see in the following for arbitrary $\xi$. Substituting in Eq.~\ref{eq:pq} we derive $q_{s}$, so the fixed points are given by
\begin{gather}
    (p_{s},q_{s},j_{zs},\phi_{s})=\left(0,\mp\frac{\gamma}{\omega}\sqrt{1-\frac{1}{f_{0x}^{2}}},-\frac{1}{f_{0x}},\pi\,\,\,\mbox{or}\,\,\ 0\right),
\end{gather}
with energy 
\begin{gather}\label{eq:ex0}
    \epsilon_{s0x}=-\frac{1}{2}\left(f_{0x}+\frac{1}{f_{0x}}\right)+\frac{\eta_{z}}{2\omega_{0}},
\end{gather}
which lies always below $\epsilon_{-}$, like in the superradiant-symmetric case. Again, the expectation value of the photon number operator becomes different from zero in the thermodynamic limit, so the phase where these points exist corresponds to a superradiant one. Its emergence is determined by $\eta_{x}$, independently of $\eta_{y}$, though. For this reason, we call it superradiant-$x$ phase. Here, the point at $j_{zs}=-1$ becomes saddle point, as it can be observed in Fig.~\ref{fig:1} (c4) to (c6) when increasing $\gamma$.  

%%%%%%%%%%%%%%%%%%%%%%%%%%%%%%%%%%%%%%%%%%
%%%%%%%%%%%%%%%%%%%%%%%%%%%%%%%%%%%%%%%%%%
\subsubsection{Superradiant-y phase}
%%%%%%%%%%%%%%%%%%%%%%%%%%%%%%%%%%%%%%%%%%
%%%%%%%%%%%%%%%%%%%%%%%%%%%%%%%%%%%%%%%%%%

This identical to the $x$ case, but in the $y$ direction. If we consider $\eta_{x}\neq\eta_{y}$ ($\Delta\eta_{zx}\neq\Delta\eta_{zy}$) and $\cos\phi_s=0$ ($\sin\phi_s=\pm1$), now $q_{s}=0$ and
\begin{gather}
    j_{zs}=-\left(\frac{\eta_{z}-\eta_{y}}{\omega_{0}}+\frac{\gamma^{2}}{\omega\omega_{0}}\right)^{-1}=-\frac{1}{f_{0y}},\,\,\,\\ \nonumber
f_{0y}=\frac{\Delta\eta_{zy}}{\omega_{0}}+f_{0+}
\end{gather}
where the critical is coupling given by
\begin{gather}\label{eq:critt0y}
\gamma_{0y}^{c}=\gamma_{0+}\sqrt{1-\frac{\eta_{z}-\eta_{y}}{\omega_{0}}}.
\end{gather}
%and $\Delta\eta_{zy}\leq\omega_{0}$. 
Substituting in Eq.~\ref{eq:pq} we can obtain $p_{s}$. The two degenerated fixed points are 
\begin{gather}
    (p_{s},q_{s},j_{zs},\phi_{s})=\left(\pm\frac{\gamma}{\omega}\sqrt{1-\frac{1}{f_{0y}^{2}}},0,-\frac{1}{f_{0y}},\pm\frac{\pi}{2}\right).
\end{gather}
Finally, their energy is
\begin{gather} \label{eq:ey0}
    \epsilon_{s0y}=-\frac{1}{2}\left(f_{0y}+\frac{1}{f_{0y}}\right)+\frac{\eta_{z}}{2\omega_{0}}.
\end{gather}
Again, $\epsilon_{s0y}\leq\epsilon_{-}$. The difference with respect to the previous superradiant phases is that now the fixed points are rotated in phase space by $\pi/2$, as can be seen from Figs.~\ref{fig:1} (c7) to (c9). As a result, it is the quadrature $q_{s}$ the one that has become zero. 

%%%%%%%%%%%%%%%%%%%%%%%%%%%%%%%%%%%%%%%%%%
%%%%%%%%%%%%%%%%%%%%%%%%%%%%%%%%%%%%%%%%%%
\subsubsection{Quantum phases in the Tavis-Cummings limit}
%%%%%%%%%%%%%%%%%%%%%%%%%%%%%%%%%%%%%%%%%%
%%%%%%%%%%%%%%%%%%%%%%%%%%%%%%%%%%%%%%%%%%
We have already seen that in the normal phase there are only two extrema in the energy surface located at $j_{zs}=\pm 1$, whereas, in the superradiant phases described above, we have found four (or a continuous set, in the symmetric case). The energy of the ground-state across the different parameter domains can be expressed in a closed form as 
\begin{gather}
    \epsilon_{0}^{\text{g.s.}}=-\frac{1}{2}\left(F_{0}+F_{0}^{-1}\right)+\frac{\eta_{z}}{2\omega_{0}}
\end{gather}
with 
\begin{gather}
    F_{0}=\left\{\begin{array}{cc}
    f_{0} & \mbox{for}\,\,\,\eta_{x}=\eta_{z},\,\,\, \mbox{and}\,\,\, \gamma\geq\gamma_{0}^{c}, 
    %\,\,\, \mbox{and}\,\,\, \Delta\eta_{zs}\leq\omega_{0}, 
    \\
    f_{0x} & \mbox{for}\,\,\,\eta_{x}\neq\eta_{y},\,\,\,\mbox{and} \,\,\, \gamma\geq\gamma_{0x}^{c},
    %\,\,\, \mbox{and}\,\,\, \Delta\eta_{zx}\leq\omega_{0}, 
    \\
    f_{0y} & \mbox{for}\,\,\,\eta_{x}\neq\eta_{y},\,\,\,\mbox{and}\,\,\, \gamma\geq\gamma_{0y}^{c},
    %\,\,\, \mbox{and}\,\,\, \Delta\eta_{zy}\leq\omega_{0}, 
    \\
    1 & \mbox{otherwise} 
    \end{array}
    \right.
\end{gather}
Let's suppose $\eta_{y}=0$, so there are no interactions in the $y$ direction. Then, as a function of $\gamma$ the system undergoes a superradiant QPT at $\gamma=\gamma_{0x}^{c}$. 
%\omega_{0}\geq\Delta\eta_{zx}$ and $f_{0x}\geq1$. 
As a result, the number of fixed points in the energy surface will change from two to four, and the minimum of the energy surface will be modified, reflecting the change in the ground-state energy associated with the QPT. The same will happen if we consider qubit-qubit interactions only in the $y$ and $z$ direction but not in the $x$ directions.

The situation is different when considering the combination of interactions in the $x$ and $y$ directions. In this case, there is a possibility that the fixed points arising from each direction simultaneously appear. Then, we will speak of a superposition of the phases. However, it is important to emphasize that even though the two phases appear superimposed, only one set of degenerate fixed points will correspond to the minimum of the energy surface, so the passage from a superradiant phase alone to a superposition of phases is not followed by a QPT (although the smooth DoS will abruptly change announcing the onset of new ESQPTs, thus the necessity to distinguish between a superradiant phase alone and a superimposed one). Here, we observe that, if $\eta_{y}\geq\eta_{x}$ ($\Delta\eta_{zx}\geq\Delta\eta_{zy}$) then we have that $\gamma_{0x}^{c}\leq\gamma_{0y}^{c}$. Without loss of generality, we will take this as the standard case for most of our expressions (otherwise, the superradiant-$x$ and $y$ phases will exchange places). This condition separates the parameter domains in three zones: the normal phase $\gamma\in[0,\gamma_{0x}^{c}]$, the superradiant-$x$ phase $\gamma\in[\gamma_{0x}^{c},\gamma_{yx}^{c}]$ and a superposition of the superradiant-$x$ and the superradiant-$y$ phases $\gamma\in[\gamma_{0y}^{c},\infty)$. Hence, depending on the values $\Delta\eta_{zx}$ and $\Delta\eta_{zx}$ the energy surface could have up to six stationary points, where only one (normal), two (superradiant), or an infinite set (superradiant-symmetric)  can correspond to the ground-state. Also, we notice that for $\Delta\eta_{zx}\geq 0$ ($\Delta\eta_{zy}\geq 0$), the system enters into the superposition of phases for every value of $\gamma$, going from four to six stationary points in the energy surface, and with a ground-state determined by the relationship between $\gamma_{0x}^{c}$ and $\gamma_{0y}^{c}$.

\begin{widetext}
\begin{figure*}
\begin{center}
\begin{tabular}{c}
\includegraphics[width=0.85\textwidth]{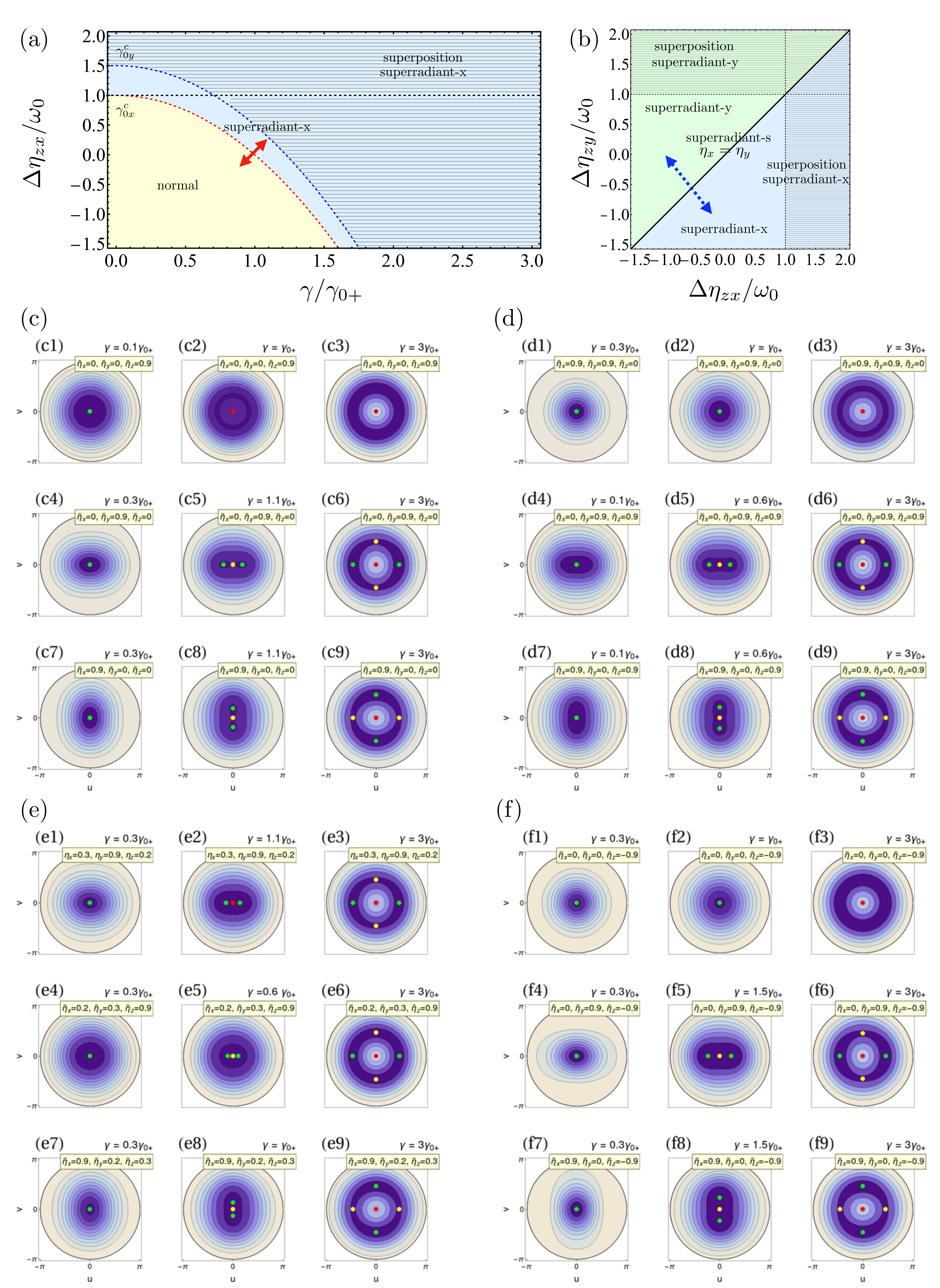}
\end{tabular}
\end{center}
\vspace{-20pt}
\caption{Quantum phases in the Tavis-Cummings limit ($\xi=0$) as a function of the Hamiltonian parameters: (a) in the $\gamma$ vs. $\Delta\eta_{zx}$ and (b) in the $\Delta\eta_{zx}$ vs  $\Delta\eta_{zy}$ spaces. The normal, superradiant-$x$ and superradiant-$y$ phases are colored in yellow, blue, and green, respectively. Filling with horizontal lines indicates the superposition of superradiant phases. The dashed red and blue curves mark the critical couplings separating the normal and superradiant phases. The dashed black curves signal the limit of validity of the superradiant phase as a function of qubit interactions. In contrast, the solid black line marks the separation between the superradiant-$x$ and $y$ phases. Red solid and blue dashed arrows indicate second and first-order QPT, respectively. (c) to (f) Contour plots of the energy surfaces in the TC limit for different parameter configurations: (c) varying the interactions in one direction while keeping the rest at zero; (d) keeping one direction to zero; (e) for arbitrary values of the interactions; (f) for negative values of the interactions. Green, red and yellow points depict minimum, maximum, and saddle fixed points on the surface, respectively. Here, $\bar{\eta}_{i}=\eta_{i}/\omega_{0}$ for all directions $i$.}
\label{fig:1} 
\end{figure*}
\end{widetext}

Thus, while in the normal phase there are only two relevant energies such that $\epsilon_{-}\leq\epsilon_{+}$, and in the superradiant-symmetric phase three $\epsilon_{s0\phi}\leq\epsilon_{-}\leq \epsilon_{+}$, in the superradiant-$x$ and $y$, we will have four: $\epsilon_{s0x}<\epsilon_{s0y}\leq\epsilon_{-}\leq \epsilon_{+}$. This will become important later in Sec.~\ref{sec:4} when discussing ESQPTs. Notice that the superradiant-symmetric phase exists for $\gamma\in[\gamma_{0}^{c},\infty)$ because $\eta_{x}=\eta_{y}$. Moreover, now we can explain the directions of the energy surface deformation in the normal phase. It turns out that if $\gamma_{0x}^{c}<\gamma_{0y}^{c}$ the deformation occurs in the $x$ direction, as it is shown in Fig.~\ref{fig:1} (c4) even if $\eta_{x}=0$. The opposite is true: if $\gamma_{0y}^{c}<\gamma_{0x}^{c}$ the deformation occurs in the $y$ direction, as shown in Fig.~\ref{fig:1} (c7). 

The ground-state energy is continuous at the critical values of the light-matter coupling $\gamma$. Nevertheless, the derivatives are discontinuous. The order of the discontinuity allows us to classify the type of quantum phase transitions the system exhibits according to Ehrenfest's classification of phase transitions. To do so, we calculate the gradient of the ground-state energy as a function of the interactions
\begin{gather}\label{eq:gradTC}
    \nabla \epsilon_{0}^{\text{g.s.}}=\left(\frac{\partial \epsilon_{0}^{\text{g.s.}}}{\partial \gamma},\frac{\partial \epsilon_{0}^{\text{g.s.}}}{\partial \eta_{x}},\frac{\partial \epsilon_{0}^{\text{g.s.}}}{\partial \eta_{y}},\frac{\partial \epsilon_{0}^{\text{g.s.}}}{\partial \eta_{z}}\right)=\\ \nonumber
    \frac{1}{2\omega_{0}}\left\{\begin{array}{cc}
     \frac{1-f^{2}_{0x}}{f^{2}_{0x}}\left(\frac{2\omega_{0}}{\gamma}f_{0x},-1,0,1\right)
       +(0,0,0,1) & \mbox{for}\,\,\,\gamma\geq\gamma_{0x}^{c}, 
       %\,\,\, \mbox{and}\,\,\, \Delta\eta_{zx}\leq\omega_{0},
    \\
      \frac{1-f^{2}_{0y}}{f^{2}_{0y}}\left(\frac{2\omega_{0}}{\gamma_{0+}}f_{0y},0,-1,1\right)+ (0,0,0,1)  & \mbox{for}\,\,\,\gamma\geq\gamma_{0y}^{c},
      %\,\,\, \mbox{and}\,\,\, \Delta\eta_{zy}\leq\omega_{0}, 
     \\
    (0,0,0,1) & \mbox{otherwise} 
    \end{array}
    \right.
\end{gather}
We need to evaluate the derivatives at three specific combinations of the parameters: $F_{0}^{c}=1$, $\Delta\eta_{zx}=\omega_{0}$ ($\Delta\eta_{zy}=\omega_{0}$) and $\Delta\eta_{zx}=\Delta\eta_{zy}$. At the critical light-matter coupling, we have $F_{0}^{c}=1$, so the ground-state energy from the normal to the superradiant phases as a function of the parameter $\gamma$ is continuous in the zeroth- and first order, and only discontinuous at the second one. Thus, a second-order phase transition occurs from normal to superradiant, as expected from the standard TC model. As a function of $\eta_{x}$ and $\eta_{y}$ there is a first-order quantum phase transition at $\Delta\eta_{x}=\Delta\eta_{y}$ because it is the border between the superradiant-$x$ and superradiant-$y$ phases, i.e., the ground-state energy goes from being described by Eq.~\ref{eq:ex0} to Eq.~\ref{eq:ey0}, passing through Eq.~\ref{eq:es0} [see Fig.~\ref{fig:1} (b)]. Finally, there are other two first-order QPTs where the system goes directly from the normal to the superradiant phase when $\Delta\eta_{zx}=\omega_{0}$ or  $\Delta\eta_{zx}=\omega_{0}$ and $\gamma=0$. 
%This was already identified in Ref.~\cite{Zhao2017}. 
This first-order phase transition %due to $z$ interactions $\eta_{z}$
was identified in Refs.~\cite{Zhao2017,Rodriguez2011}. Here we have found a generalization discovering that the relevant parameter is not just $\eta_{z}$, but $\Delta\eta_{zx}$ (or $\Delta\eta_{zx}$) instead. This is not surprising because given that the pseudospin length is conserved a single direction can be expressed in terms of the others, i.e., $\hat{J}^{2}_{z}=j(j+1)\hat{\mathbb{I}}-\hat{J}^{2}_{x}-\hat{J}^{2}_{y}$, so the qubit interacting Hamiltonian can be written as
\begin{gather}
\hat{H}_{qq}=-\frac{1}{N}\left(\Delta\eta_{zx}\hat{J}_{x}^{2}+\Delta\eta_{zy}\hat{J}_{y}^{2}\right)+\frac{1}{N}\eta_{z}j(j+1)\hat{\mathbb{I}}
\end{gather}
Hence, the role of $x$ or $y$ interactions is to shift the critical coupling of the superradiant QPT and to shift the value of the interactions where the first-order phase transition emerges. In Fig.~\ref{fig:1} (a) and we show the quantum phases in the $\Delta\eta_{zx}$ versus $\gamma$ space. There, we have included an artificial shift $\omega_{0}/2$ between $\eta_{x}$ and  $\eta_{y}$ to exhibit the onset of the superradiant-$x$ followed by a superposition of phases as the light-matter coupling increases, otherwise, the phases overlap in the diagram (because of the symmetry). In Fig.~\ref{fig:2} (b) we show the quantum phases in the $\Delta\eta_{zy}$ versus $\Delta\eta_{zx}$ space. This diagram depends on the value of $\gamma$; here, we choose an illustrative value to always be in the superradiant phases. 

Further information about the system can be gained by studying the energy surfaces as they are shown in Fig.~\ref{fig:1} (c)-(f). We employ representative values of the interaction strengths in each direction to explore and better show the evolution of the surface as the interacting parameters change. In the first row of Fig.~\ref{fig:1} (c1)-(c3) we present a configuration $\eta_{x}=\eta_{y}$, where we expect symmetry in the system. Increasing the light-matter coupling makes the surface go from a spherical well with a minimum at the center to the Mexican hat potential characteristic of the superradiant-symmetric phase, as shown in Fig.~\ref{fig:1} (c1). Next, the symmetry of the surface in both the normal and superradiant phases is broken once we increase the interactions in a given direction other than $z$. In the middle row of Fig.~\ref{fig:1} (c) we make $\eta_{y}=0.9\omega_{0}$ and $\eta_{x}=\eta_{z}=0$. As mentioned before, here we identify a stretching of the energy surface in the $x$ direction (because $\epsilon_{s0x}<\epsilon_{s0y}$). In Fig.~\ref{fig:1} (c4) we select a coupling $\gamma<\gamma^{c}_{0x}$ that locate us within the normal phase. Eventually, increasing $\gamma$ leads us to the superradiant-$x$ phase with two minima and a saddle point, as shown in Fig.~\ref{fig:1} (c5). Then, in Fig.~\ref{fig:1} (c6), we enter a regime of superposition between the two superradiant phases. Now, there are two minima points, two saddle points, and a maximum in the center. The situation is the same in the third row. From Fig.~\ref{fig:1} (c7)-(c9), now the stretching of the energy surface is in the y-direction, and we will recover the phenomenology of the last case but rotated in $\pi/2$, characteristic of the superradiant-$y$ phase. Here, for larger $\gamma$, the dominant phase is the superradiant-$y$ as $\epsilon_{s0y}<\epsilon_{s0x}$. The reasoning is the same in the other examples we show in Figs.~\ref{fig:1} (d)-(f). In Fig.~\ref{fig:1} (d), we fix one of the directions to zero and tune the other two equal to 1. In the first two rows, we show the evolution of the surface as we increase $\gamma$ while making $\eta_{x}=\eta_{z}=0.9$ ($\eta_{y}=\eta_{z}=0.9$). The situation becomes similar to that of Fig.~\ref{fig:1} because the relevant parameters to determine the phases are $\Delta\eta_{zx}$ and $\Delta\eta_{zy}$. In Fig.~\ref{fig:1} (e) we choose a different combination of values for the interactions in each direction. Finally, in Fig.~\ref{fig:1} (f) we select negative values of the interactions. Given a set of interacting parameters $\eta_{i}$ we can identify how the energy surface will be deformed, and increasing $\gamma$ leads us always to the superposition of phases $x$-$y$. In each figure, we will have different couplings $\gamma$ to highlight the system's phase.

Next, we will explore the other limit of the Hamiltonian, when the non-resonant terms are completely included, i.e., the Dicke limit.

%%%%%%%%%%%%%%%%%%%%%%%%%%%%%%%%%%%%%%%%%%
%%%%%%%%%%%%%%%%%%%%%%%%%%%%%%%%%%%%%%%%%%
%%%%%%%%%%%%%%%%%%%%%%%%%%%%%%%%%%%%%%%%%%
%%%%%%%%%%%%%%%%%%%%%%%%%%%%%%%%%%%%%%%%%%

%%%%%%%%%%%%%%%%%%%%%%%%%%%%%%%%%%%%%%%%%%
%%%%%%%%%%%%%%%%%%%%%%%%%%%%%%%%%%%%%%%%%%
\subsection{Dicke limit}
%%%%%%%%%%%%%%%%%%%%%%%%%%%%%%%%%%%%%%%%%%
%%%%%%%%%%%%%%%%%%%%%%%%%%%%%%%%%%%%%%%%%%

We now fully include the counter-rotating terms, i.e., we set $\xi=1$. This corresponds to the Dicke limit, meaning we obtain a Dicke Hamiltonian plus the qubit-qubit interactions. It is given by
\begin{gather}
    H_{cl}^{(1)}=\frac{\omega}{2}(q^{2}+p^{2})+j_{z}\left(\omega_{0}+\frac{\eta_{z}j_{z}}{2}\right)\\\nonumber+\frac{1}{2}\left(1-j^{2}_{z}\right)\left(\eta_{x}\cos^{2}\phi+\eta_{y}\sin^{2}\phi\right)\\ \nonumber
    +2\gamma\sqrt{1-j^{2}_{z}}q\cos\phi.
\end{gather}
Eqs.~\ref{eq:c1} and~\ref{eq:c2} become
\begin{gather} \label{eq:c1d}
    \omega_{0}+j_{zs}\left(\eta_{z}-\left(\eta_{x}\cos^{2}\phi_{s}+\eta_{y}\sin^{2}\phi_{s}\right)+\frac{4\gamma^{2}}{\omega}\cos^{2}\phi_{s}\right)=0,\\
    (1-j_{zs}^{2})\left[(\eta_x-\eta_y)-\frac{4\gamma^{2}}{\omega}\right]\cos\phi_{s}\sin\phi_{s}=0.
\end{gather}
Just as in the case of the TC limit, we obtain five possibilities when searching stationary points of the energy surface. The first four are identical: $j_{zs}=\pm1$, $\cos\phi_{s}=0$ ($\sin\phi_{s}=\pm1$), $\sin\phi_{s}=0$ ($\cos\phi_{s}=\pm1$). However, the last one contains a different condition for the parameters, given by $4\gamma^{2}/\omega=\eta_x-\eta_y$. We immediately notice that the symmetry that was found in the TC limit ($\xi=0$) is now always broken because the relation $\eta_{x}=\eta_{y}$ does not lead to the existence of stationary points anymore. This is an expected result given that the standard Dicke model is non-integrable~\cite{Bastarrachea2015}. Also, from Hamilton equations, we observe that $p_{s}$ is always zero in this limit.

We have a normal phase in the Dicke limit too, where the fixed point at $j_{zs}=-1$ is an absolute minimum corresponding to the ground-state, and the point at $j_{zs}=+1$ is an absolute maximum. Both points have an energy given by $\epsilon_{\pm}=\pm 1 +\eta_{z}/2\omega_{0}$ and the normal phase will be deformed if the interactions are privileged in either the $x$ or $y$ as it happened in the $\xi=0$ limit. This case is shown in Fig.~\ref{fig:2} (c1), (c4, and (c7), where we encounter the same deformations as in the TC limit. Next, we will explore the other phases appearing for $\xi=1$. 

%%%%%%%%%%%%%%%%%%%%%%%%%%%%%%%%%%%%%%%%%%
%%%%%%%%%%%%%%%%%%%%%%%%%%%%%%%%%%%%%%%%%%
\subsubsection{Superradiant-x phase}
%%%%%%%%%%%%%%%%%%%%%%%%%%%%%%%%%%%%%%%%%%
%%%%%%%%%%%%%%%%%%%%%%%%%%%%%%%%%%%%%%%%%%

When we have $\cos\phi_{s}=\pm 1$ ($\sin\phi_{s}=0$) we encounter the same superradiant phase as in the $\xi=0$ limit, but with a modified critical coupling, as it would be expected~\cite{Bastarrachea2014a}. Here, the value of the collective atomic variable becomes
\begin{gather}
    j_{zs}=-\frac{1}{f_{1x}},\,\,\,
    f_{1x}=\frac{\Delta\eta_{zx}}{\omega_{0}}+f_{1+}
\end{gather}
where $f_{1x}=\gamma^{2}/\gamma_{1+}^{2}$ and $\gamma_{1+}=\sqrt{\omega\omega_{0}}/2$ is the critical coupling of the standard Dicke model. Using Eq.~\ref{eq:pq} one gets the two degenerate minima typical of the superradiant phase 
\begin{gather}
     \left(p_{s},q_{s},jz_{s},\phi_{s}\right)=\left(0,\pm\frac{2\gamma}{\omega}\sqrt{1-\frac{1}{f_{1x}^{2}}},-\frac{1}{f_{1x}},\,0\,\,\,\text{or}\,\,\,\pi\right)
\end{gather}
The energy associated to these points is
\begin{gather}
    \epsilon_{s1x}=-\frac{1}{2}\left(f_{1x}+\frac{1}{f_{1x}}\right)+\frac{\eta_{z}}{2\omega_{0}}.
\end{gather}
An almost identical result as that of the TC-like superradiant-$x$ phase. The ground-state energy of the system is always below $\epsilon_{-}$; thus, the energy of these points corresponds to the ground-state energy. This phase also is valid for values of the light-matter coupling $\gamma\geq\gamma_{1x}^{c}$, where
\begin{gather}
    \gamma_{1x}^{c}=\gamma_{1+}\sqrt{1-\frac{\Delta\eta_{zx}}{\omega_{0}}}.
\end{gather}
%and $\Delta\eta_{zx}\leq\omega_{0}$. 
Suppose the interactions $\eta_{x,z}$ vanish. In that case, we recover the standard result for the Dicke model, again, we observe that the role of the interactions is to shift the critical coupling. This effect has been observed before in several previous works for interactions in the $x$~\cite{Jaako2016,DeBernardis2018} and $z$~\cite{Chen2008,Chen2010,YiXiang2013,Zhao2017} directions, and a combination of $x$ and $y$ directions~\cite{Chen2006}. Not only in the case of the Dicke limit but for arbitrary $\xi$, the critical coupling to attain superradiance can become zero with a suitable choice of the interactions in the $z$ and $x$ ($y$) directions, given that the relevant parameter is the difference between the $\Delta\eta_{zx}$ ($\Delta\eta_{zy}$) relative interactions, as we have seen before.

A specific feature of the $\xi=1$ limit is that breaking the rotational symmetry leads to the exclusion of the superradiant-$y$ phase. Instead, we have the deformed phase, as we will immediately explain. 

%%%%%%%%%%%%%%%%%%%%%%%%%%%%%%%%%%%%%%%%%%
%%%%%%%%%%%%%%%%%%%%%%%%%%%%%%%%%%%%%%%%%%
\subsubsection{Deformed phase}
%%%%%%%%%%%%%%%%%%%%%%%%%%%%%%%%%%%%%%%%%%
%%%%%%%%%%%%%%%%%%%%%%%%%%%%%%%%%%%%%%%%%%

Instead of the superradiant-$y$ phase, there is a new quantum phase emerging from the interactions. From the condition where $\cos\phi_{s}\pm1$ ($\sin\phi_{s}=0$), we get
\begin{gather}
    j_{zs}=-\frac{\omega_0}{\Delta\eta_{zy}}=-\frac{1}{f_{1y}},
\end{gather}
This leads to two new fixed points
\begin{gather}
         \left(p_{s},q_{s},jz_{s},\phi_{s}\right)=\left(0,0,-\frac{1}{f_{1y}},\pm\frac{\pi}{2}\right),
\end{gather}
whose energy is given by
\begin{gather}
    \epsilon_{s1y}=-\frac{1}{2 f_{1y}}+\frac{\eta_{y}}{2\omega_{0}},
\end{gather}
However, if we add and subtract $\eta_{z}/2\omega_{0}$ it reads
\begin{gather}
    \epsilon_{s1y}=-\frac{1}{2}\left(f_{1y}+\frac{1}{f_{1y}}\right)+\frac{\eta_{z}}{2\omega_{0}},
\end{gather}
The main difference with previous cases (the superradiant phases) is that $f_{1y}$ is independent of $\gamma$, so this phase exists for every value of the light-matter coupling given that $\Delta\eta_{zy}\geq \omega_{0}$. This phase was identified before in Ref.~\cite{Rodriguez2018} in the absence of $\eta_{y}$. It is characterized by the two degenerate fixed points whose orientation in the atomic angle $\phi$ is rotated by $\pi/2$ with respect to the superradiant-$x$ phase's fixed points (a result inherited from the superradiant-$y$ phase). There, the photon number's expectation value becomes zero because $q_{s}=p_{s}=0$. Also, we have that because $|f_{1y}|\leq 1$, $\epsilon_{s1y}\leq\epsilon_{-}<\epsilon_{+}$. Then, they mark the ground-state energy, and the point at $j_{zs}=-1$ becomes a saddle point, as it happens in the usual superradiant phases. However, because it is neither a normal nor a superradiant phase, we deem it as a {\it deformed} phase, although one could name it a subrradiant phase. The energy surfaces in this phase are shown in Figs.~\ref{fig:2} (e7) and (f7).   

Finally, we study the last condition for stationary points, given by the parameter relation  
\begin{gather}\label{eq:cd1}
  \frac{\eta_{x}-\eta_{y}}{\omega_{0}}=\frac{\Delta\eta_{zy}-\Delta\eta_{zx}}{\omega_{0}}=\frac{\gamma^{2}}{\gamma_{1+}^{2}}=f_{1+}
\end{gather}
Because we can write Eq.~\ref{eq:c1d} as
\begin{gather}
    \omega_{0}+j_{zs}\left[\eta_{z}-\eta_{y}-\cos^{2}\phi\left(\eta_{y}-\eta_{x}+\frac{4\gamma^{2}}{\omega}\right)\right]=0,
\end{gather}
it is clear that, when applying the condition in Eq.~\ref{eq:cd1}, the factor multiplying the cosine function vanishes so we get a stationary point. Here, we find the following value of $j_{zs}$ with energy
\begin{gather}
   \left(p_{s},q_{s},jz_{s},\phi_{s}\right)=\left(0,0,-\frac{1}{f_{1y}},\pm\frac{\pi}{2}\right),\\ \nonumber
   \epsilon_{s1y}=-\frac{1}{2}\left(f_{1y}+\frac{1}{f_{1y}}\right)+\frac{\eta_{z}}{2\omega_{0}},
\end{gather}
i.e., the stationary points from the deformed phase. Nevertheless, because we can write Eq.~\ref{eq:cd1} as
\begin{gather}
\frac{\Delta\eta_{zy}}{\omega_{0}}=\frac{\Delta\eta_{zx}}{\omega_{0}}+f_{1+}.
\end{gather}
Therefore, $\epsilon_{s1x}=\epsilon_{s1y}$ and the stationary points coincide with those of the superradiant-$x$ phase. This means that Eqs.~\ref{eq:cd1} marks the frontier between the superradiant-$x$, the deformed phases, and the normal phase (for $f_{1y}=1$), a similar result to $\eta_{x}=\eta_{y}$ for $\xi=0$.

%%%%%%%%%%%%%%%%%%%%%%%%%%%%%%%%%%%%%%%%%%
%%%%%%%%%%%%%%%%%%%%%%%%%%%%%%%%%%%%%%%%%%
\subsubsection{Quantum phases in the Dicke limit}
%%%%%%%%%%%%%%%%%%%%%%%%%%%%%%%%%%%%%%%%%%
%%%%%%%%%%%%%%%%%%%%%%%%%%%%%%%%%%%%%%%%%%

The existence of the deformed phase changes the quantum phase diagram in the Dicke limit. In the TC, there is a chance for a superposition of the two superradiant phases. We recall that in this case, only one set of stationary points becomes the minimum, either those from the superradiant-$x$ or those from the superradiant-$y$ phases. Instead, there is a stricter separation between phases because the deformed phase appears for $\Delta\eta_{zy}\geq\omega_{0}$. Then, we have only two stationary points at the normal phase and only four for both the superradiant-$x$ and deformed phases. We can write the ground-state energy in closed form too:
\begin{gather}
    \epsilon_{1}^{\text{g.s.}}=-\frac{1}{2}\left(F_{1}+F_{1}^{-1}\right)+\frac{\eta_{z}}{2\omega_{0}}
\end{gather}
with 
\begin{gather}
    F_{1}=\left\{\begin{array}{cc}
    f_{1x} & \mbox{for}\,\,\,\gamma\geq\gamma_{1x}^{c}
    \,\,\, \mbox{and}\,\,\, \Delta\eta_{zx}\leq\omega_{0}, 
    \\
    f_{1y} & \mbox{for}\,\,\, \Delta\eta_{zy}\geq\omega_{0}, 
    \\
    1 & \mbox{otherwise} 
    \end{array}
    \right.
\end{gather}
We also remind that $f_{1y}$ is independent from $\gamma$. Again, if we suppose $\eta_{y}=\eta_{z}=0$, the ground-state evolves as a function of $\gamma$ from the normal $\gamma\in[0,\gamma_{1x}^{c}]$ to the superradiant-$x$ phase $\gamma\in[\gamma_{1x}^{c},\infty)$. Then, the situation remains the same for $\Delta\eta_{zy}\neq0$, but the deformed phase emerges above $\omega_{0}$. Similarly to the TC limit we will have the following energy intervals $\epsilon_{s1x}<\epsilon_{s1y}<\epsilon_{-}<\epsilon_{+}$ in the superradiant-$x$ phase and $\epsilon_{s1y}<\epsilon_{-}<\epsilon_{+}$ in the deformed phase. 

Next, we obtain the gradient of the ground-state energy as a function of the interactions, just like we did in the TC regime:
\begin{widetext}
\begin{gather}\label{eq:gradDicke}
    \nabla \epsilon_{1}^{\text{g.s.}}=  
    \frac{1}{2\omega_{0}}\left\{\begin{array}{cc} 
    \frac{1-f_{1x}^{2}}{f^{2}_{1x}}\left(\frac{2\omega_{0}}{\gamma}f_{1x},-1,0,1\right)+(0,0,0,1) & \mbox{for}\,\,\,\gamma\geq\gamma_{1x}^{c}, 
    \,\,\, \mbox{and}\,\,\, \Delta\eta_{zx}\leq\omega_{0}, 
    \\
    \frac{1-f^{2}_{1y}}{f^{2}_{1y}}\left(0,0,-1,1\right)+ (0,0,0,1)  &  \mbox{for}\,\,\,\Delta\eta_{zy}\geq\omega_{0},  \\
   (0,0,0,1) & \mbox{otherwise} 
    \end{array}
    \right., 
\end{gather}
\end{widetext}

\begin{widetext}
\begin{figure*}
\begin{center}
\begin{tabular}{c}
\includegraphics[width=0.8\textwidth]{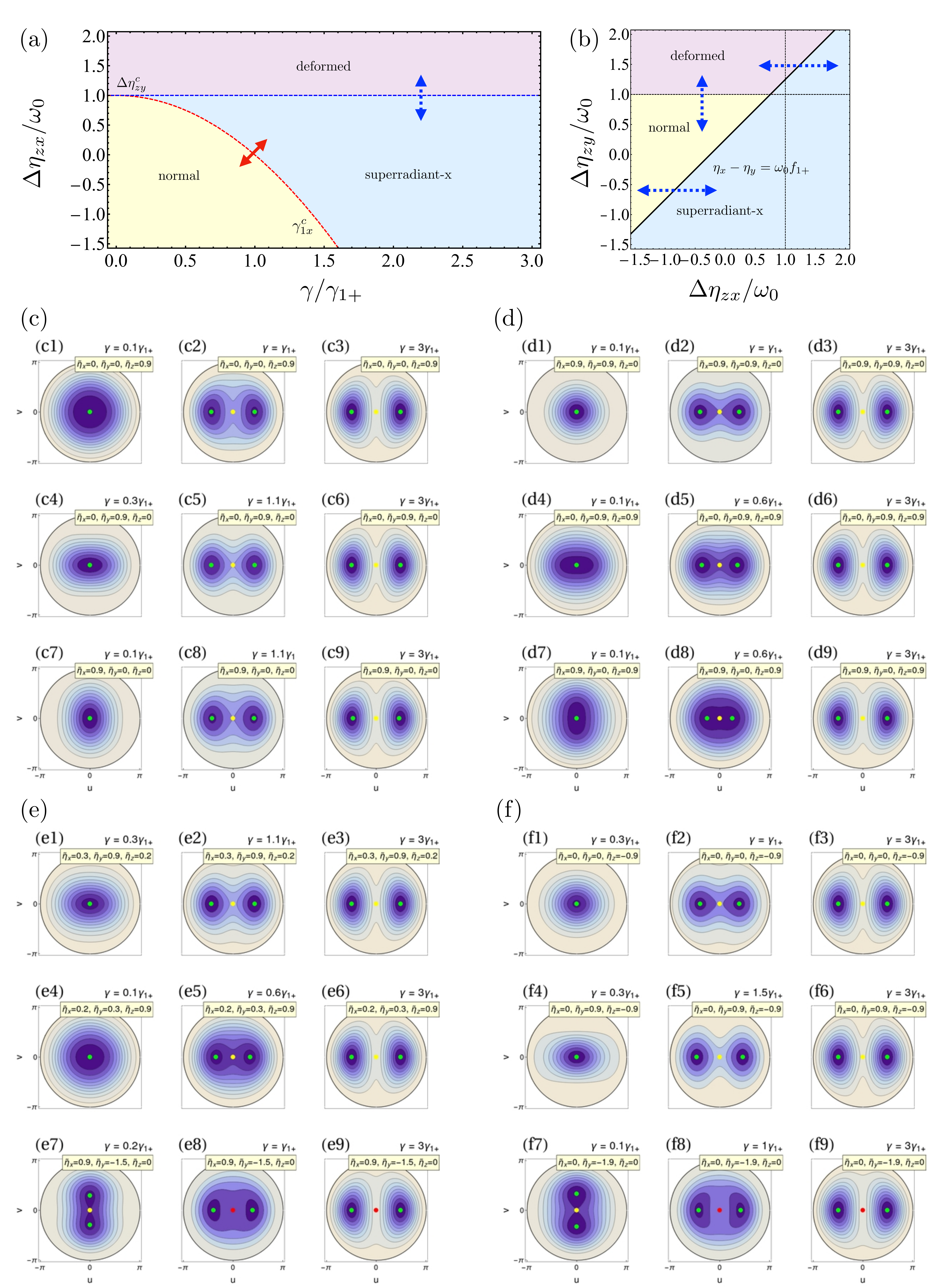}
\end{tabular}
\end{center}
\vspace{-20pt}
\caption{The same as in Fig.~\ref{fig:1}, but for the Dicke limit ($\xi=1$). The deformed phase is indicated in purple. The quantum phase diagram in Fig.~\ref{fig:2} (b) is for $\gamma=0.5\gamma_{1+}$. Likewise, the thick solid black line in (b) is calculated for given value of $\gamma$ such $f_{1+}=0.5$.}
\label{fig:2} 
\end{figure*}
\end{widetext}

There are three sets of parameter values to look for the presence of QPT. First, $F_{1}^{c}=1$, where we get a generalization of well-known second-order Dicke QPT from the normal to the superradiant-$x$ phase. Next, at $\Delta\eta_{zy}^{c}=\omega_{0}$ we get a first-order QPT from the normal and superradiant-$x$ phases to the deformed one as a function of $\Delta\eta_{zx}$ or $\Delta\eta_{zy}$, recovering the results in Ref.~\cite{Rodriguez2018} when $\eta_{x}=\eta_{y}=0$. Finally, at  $\Delta\eta_{zy}-\Delta\eta_{zx}=\omega_{0}f_{1+}$, we have the corresponding behavior we found for the TC limit at $\eta_{x}=\eta_{y}$: there is a first-order QPT signaling the border between the $x$ and $y$ sides. In Figs.~\ref{fig:2} (a)-(b), we show the different phases in the system as a function of the Hamiltonian parameters. The presence of first-order phase transitions in the Dicke model due to qubit-qubit interactions were predicted before considering interactions in the $y$ direction~\cite{Lee2004}, and later in the $z$~\cite{Chen2008,Robles2015,Zhao2017}, and are inherited from the LMG model as well. 

The energy surfaces for the same parameter configurations we employed in the TC limit are shown in Fig.~\ref{fig:2} (c) to (f). In some cases, we slightly change the value of the interactions to highlight the phase in which the system is located. We observe the onset of the fixed points and what kind of extreme point they correspond to (minimum, maximum, saddle point) in each phase as a function of the interacting parameters. For $\eta_{x}=\eta_{y}=0$ [Fig.~\ref{fig:2} (c1)-(c3)] the system goes from the normal to the superradiant-$x$ case, but the normal phase is not deformed. As usual, changing the balance between $\eta_{x}$ and $\eta_{y}$ breaks the overall symmetry of the normal phase but, in this case, leaves unaffected the superradiant phase. In the Dicke limit, we have only a restriction for $\gamma$ given by $\gamma^{c}_{1x}$, as the deformed phase is independent of the light-matter coupling. As a direct consequence, the deformation tends to stretch in the horizontal direction, but we have cases as those in Fig.~\ref{fig:2} (c7) and (d7), where the energy surface in the normal phase is deformed in the $y$ direction. We can observe that for most sets of parameters, the situation is similar to that of Figs.~\ref{fig:2} (c4)-(c6). In (c4) $\gamma<\gamma^{c}_{1x}$, the system is in the normal phase. Increasing the parameter $\gamma$ makes the system enter the superradiant-$x$ phase where the minima are in the horizontal direction, and a saddle point appears. We can identify the appearance of the deformed phase in  (e7) and (f7), where the minima emerge in the vertical direction, and the point at the center becomes a saddle point. Finally, we stress that there is no situation where more than four stationary points appear, contrasting with the TC limit.  

%%%%%%%%%%%%%%%%%%%%%%%%%%%%%%%%%%%%%%%%%%
%%%%%%%%%%%%%%%%%%%%%%%%%%%%%%%%%%%%%%%%%%
%%%%%%%%%%%%%%%%%%%%%%%%%%%%%%%%%%%%%%%%%%
%%%%%%%%%%%%%%%%%%%%%%%%%%%%%%%%%%%%%%%%%%

%%%%%%%%%%%%%%%%%%%%%%%%%%%%%%%%%%%%%%%%%%
%%%%%%%%%%%%%%%%%%%%%%%%%%%%%%%%%%%%%%%%%%
\subsection{Arbitrary coupling}
%%%%%%%%%%%%%%%%%%%%%%%%%%%%%%%%%%%%%%%%%%
%%%%%%%%%%%%%%%%%%%%%%%%%%%%%%%%%%%%%%%%%%

We are ready to treat the general case, i.e., when $\xi\in(0,1)$. Here we expect to find effects similar to those in the TC and Dicke limits. In the absence of qubit interactions, the main difference with those limits is the presence of two different superradiant domains that can coexist for some values of the light-matter coupling. Hence, three phases separated by the critical values of the light-matter coupling appear $\gamma_{\xi\pm}=\sqrt{\omega\omega_{0}}/(1\pm\xi)$, such that $\gamma_{\xi +}<\gamma_{\xi-}$~\cite{Bastarrachea2016,Kloc2017,Stransky2017b,Cejnar2021}. Given that $\gamma_{\xi+}<\gamma_{\xi-}$, in the interval $\gamma_{\xi+}<\gamma<\gamma_{\xi-}$ one finds the standard effect of the Dicke model that we will call here superradiance-$(+)$. Instead, for $\gamma_{\xi-}<\gamma$, there are two additional stationary points whose nature is that of saddle points that are attributed to a superradiance-$(-)$ effect. However, in this superposition of superradiant phases, the fixed points from the superradiance-$(+)$ are the minima, so there is no QPT~\cite{Bastarrachea2016}. We immediately notice the similarity with what we have observed in the TC model in the presence of qubit interaction. A result that will take importance later when we explore the energy domains. If we take $\xi\rightarrow 0$, then $\gamma_{0+}=\gamma_{0-}$, and for the TC limit, the two superradiant phases become one. We can anticipate that the presence of the interactions $\eta_{x}$ and $\eta_{y}$ creates a similar result and produces two domains separated by the two critical couplings we have already described $\gamma_{0x}^{c}$ and $\gamma_{0y}^{c}$. On the other hand, when $\xi\rightarrow 1$ the critical coupling $\gamma_{1-}\rightarrow \infty$, the new superradiant-$(-)$ phase is pushed to larger values of the coupling until it vanishes. Hence, in the Dicke model, there is only one superradiant phase. In our case, this effect has been reflected on the onset of the deformed phase. As we will discuss below, by including the qubit-qubit interactions, we get a similar result for arbitrary $\xi$, where the superradiant-$(+)$ [($(-)$)] phase is modified by interactions in $x$ ($y$) directions.

Once more, we get five conditions for fixed points from Eqs.~\ref{eq:c1} and~\ref{eq:c2}: $j_{zs}=\pm 1$, $\cos\phi_{s}=\pm1$ ($\sin\phi_{s}=0$), $\sin\phi_{s}=\pm1$ ($\cos\phi_{s}=0$), and the special parameter relationship that now takes the form:
\begin{equation} \label{eq:xic5}
\frac{\eta_{x}-\eta_{y}}{\omega_{0}}=\frac{\Delta\eta_{zy}-\Delta\eta_{zx}}{\omega_{0}}=\frac{\gamma^{2}}{\gamma^{2}_{\xi+}}-\frac{\gamma^{2}}{\gamma^{2}_{\xi-}}=f_{\xi+}-f_{\xi-},
\end{equation}
$j_{zs}=\pm 1$ leads to the two stationary points that mark the absolute minimum in the normal phase and the absolute maximum. Thus, we have again a (deformed) normal phase, as in the two previous cases, Next, we will recover the most general superradiant phases.

%%%%%%%%%%%%%%%%%%%%%%%%%%%%%%%%%%%%%%%%%%
\subsubsection{Superradiant-x and y phases}
%%%%%%%%%%%%%%%%%%%%%%%%%%%%%%%%%%%%%%%%%%
If we evaluate the condition $\cos\phi_{s}=\pm 1$ ($\sin\phi_{s}=0$) we obtain the two degenerate stationary points corresponding to the superradiant ground-state given by 
\begin{gather}
        \left(p_{s},q_{s},jz_{s},\phi_{s}\right)=
         \left(0,\mp\frac{2\gamma}{\omega}\sqrt{1-\frac{1}{f_{\xi x}^{2}}},-\frac{1}{f_{\xi x}},0\,\,\,\mbox{or}\,\,\,\pi\right),
\end{gather}
where now
\begin{gather}
f_{\xi x}=\frac{\Delta\eta_{zx}}{\omega_{0}}+f_{\xi +}.
\end{gather}
At these points, the energy surface becomes
\begin{gather}
    \epsilon_{s\xi x}=-\frac{1}{2}\left(f_{\xi x}+\frac{1}{f_{\xi x}}\right)+\frac{\eta_{z}}{2\omega_{0}}
\end{gather}
and they exist for $\gamma\geq\gamma_{\xi x}^{c}$ 
%and $\Delta\eta_{zx}\leq\omega_{0}$ 
with
\begin{gather}
\gamma_{\xi x}^{c}=\gamma_{\xi +} \sqrt{1-\frac{\Delta\eta_{zx}}{\omega_{0}}}.
\end{gather}
Symmetrically, if we opt for the case $\sin\phi_{s}=\pm 1$ ($\cos\phi_{s}=0$) the stationary points are rotated by $\pi/2$, as expected,
\begin{gather}
        \left(p_{s},q_{s},jz_{s},\phi_{s}\right)=
         \left(\pm\frac{2\gamma}{\omega}\sqrt{1-\frac{1}{f_{\xi y}^{2}}},0,-\frac{1}{f_{\xi y}},\pm\frac{\pi}{2}\right),
\end{gather}
where
\begin{gather}
f_{\xi y}=\frac{\Delta\eta_{zy}}{\omega_{0}}+f_{\xi -}.
\end{gather}
In the same way, these points appear only for $\gamma\geq\gamma_{\xi y}$ with
\begin{gather}
\gamma_{\xi y}^{c}=\gamma_{\xi -} \sqrt{1-\frac{\Delta\eta_{zy}}{\omega_{0}}}.
\end{gather}
%and $\Delta\eta_{zy}\leq\omega_{0}$. 
Their energy is
\begin{gather}
    \epsilon_{s\xi y}=-\frac{1}{2}\left(f_{\xi y}+\frac{1}{f_{\xi y}}\right)+\frac{\eta_{z}}{2\omega_{0}}.
\end{gather}
As anticipated, these phases correspond to a generalization of the superradiant$-x$ and $y$ we have found in the TC and Dicke limits. Similar to the case of the Dicke model, the condition in Eq.~\ref{eq:xic5} lies at the border between the superradiant phases, as it is exhibited in Fig.~\ref{fig:3} (b).  

%%%%%%%%%%%%%%%%%%%%%%%%%%%%%%%%%%%%%%%%%%
%%%%%%%%%%%%%%%%%%%%%%%%%%%%%%%%%%%%%%%%%%
\subsubsection{Quantum phases for arbitrary coupling}
%%%%%%%%%%%%%%%%%%%%%%%%%%%%%%%%%%%%%%%%%%
%%%%%%%%%%%%%%%%%%%%%%%%%%%%%%%%%%%%%%%%%%

For arbitrary $\xi$ several of the results we have found before for $\xi=0,1$ stand. The major difference was the existence of the deformed phase in the Dicke. In fact, the phase diagram for $\xi\in(0,1)$ is very similar to the one for the TC limit, except for the absence of symmetry (for intermediate $\xi$ the Hamiltonian is non-integrable) and for the dependence of the border between the superradiant $x$ and $y$ phases given by Eq.~\ref{eq:xic5} that now is explicitly in terms of $\xi$ and $\gamma$. 

What is truly new about the intermediate $\xi$ case is the correspondence between the interactions in the $x$ direction and the superradiant-$(+)$ phase and those in the $y$ direction and the superradiant-$(-)$ phase. This effect explains our previous findings. For $\xi=0$, both phases are completely analogous except that one depends on $\eta_{x}$ and the other on $\eta_{y}$. Instead, for $\xi=1$ the superradiant-$y$ phase vanishes completely because $\gamma_{\xi-}$ goes to infinity. ($f_{\xi y}\rightarrow \Delta\eta_{zy}/\omega_{0}$). Therefore, in the Dicke limit, the superradiant-$y$ transforms into the deformed phase. The correspondence is explicit once we recognize the dependence of the light-matter critical couplings on the interactions in the $x$ and $y$ direction.

\begin{widetext}
\begin{figure*}
\begin{center}
\begin{tabular}{c}
\includegraphics[width=0.8\textwidth]{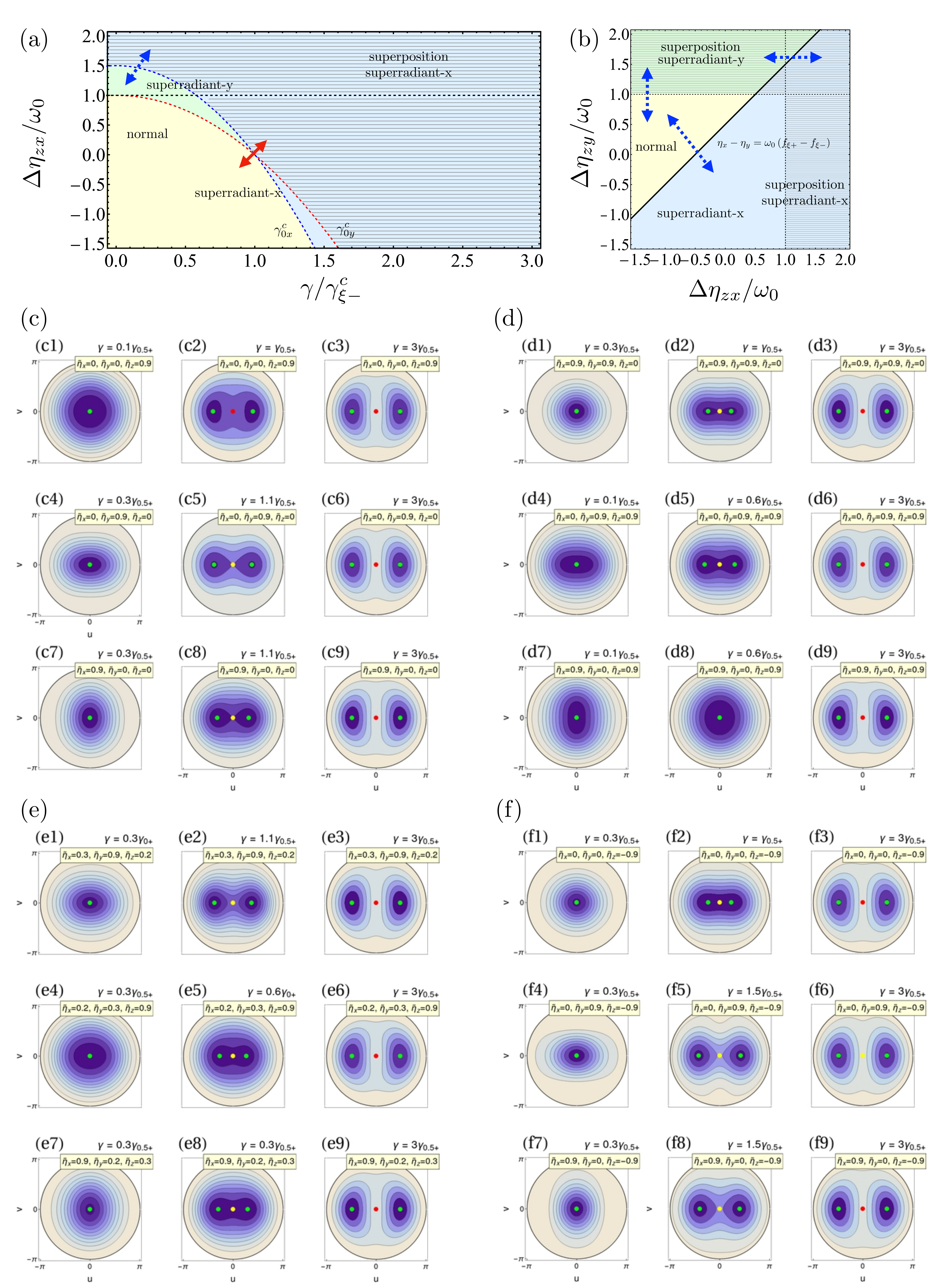}
\end{tabular}
\end{center}
\vspace{-20pt}
\caption{The same as in Fig.~\ref{fig:1}, but for an arbitrary coupling set at $\xi=0.1$ in (a) and at $\xi=0.5$ from (b) to (f). We have selected a shift $\eta_{x}-\eta_{y}=\omega_{0}/2$. The thick solid black line in (b) is calculated for $f_{1+}=0.5$. Here, $\bar{\eta_{i}}=\eta_{i}/\omega_{0}$ for all directions $i$.}
\label{fig:3} 
\end{figure*}
\end{widetext}

Surprisingly, qubit interactions can shift the order in which the two superradiant phases $x$ and $y$ occur, a situation we have already encountered in the TC limit but excluded in the standard case, i.e., for the superradiant-$(+)$ and $(-)$ phases. For $\eta_{x}=\eta_{y}=0$ it holds that $\gamma_{\xi x}^{c}\leq\gamma_{\xi y}^{c}$ (because $\gamma_{\xi +}^{c}\leq\gamma_{\xi -}^{c}$). However, if we tune $\eta_{x}$ and $\eta_{y}$ independently we can make that $\gamma_{\xi y}^{c}<\gamma_{\xi x}^{c}$. The condition to invert the order of the two critical couplings occurs for the set of parameters where 
\begin{gather}
\Delta\eta_{zy}=\frac{\gamma_{\xi+}^{2}}{\gamma_{\xi -}^{2}}\Delta\eta_{zx}-\omega_{0}\left[1-\frac{\gamma_{\xi +}^{2}}{\gamma_{\xi -}^{2}}\right]
%\frac{\eta_{y}}{\omega_{0}}-\frac{\gamma_{\xi +}^{2}}{\gamma_{\xi -}^{2}}\frac{\eta_{x}}{\omega_{0}}=\left(\frac{\gamma_{\xi +}^{2}}{\gamma_{\xi -}^{2}}-1\right)\left(1-\frac{\eta_{z}}{\omega_{0}}\right)
\end{gather}
For $\xi=0$ this condition becomes $\eta_{x}=\eta_{y}$, because $\gamma_{0 +}^{c}=\gamma_{0 -}^{c}$. It is the same special parameter relation we find earlier leading to the superradiant-symmetric phase. For $\xi=1$ we have that $\left(\gamma_{1-}^{c}\right)^{-1}=0$. Then, the condition is now $\Delta\eta_{zy}=\omega_{0}$, the limiting condition where the deformed phase emerges. This effect is shown in Fig.~\ref{fig:3} (a) for $\xi=0.1$, where we have introduced an artificial shift $\Delta\eta_{zy}=\Delta\eta_{zx}+\omega_{0}/2$ (the equivalent of what we have done in Fig.~\ref{fig:1}) to exhibit that the two curves of $\gamma_{\xi x}^{c}$ and $\gamma_{\xi y}^{c}$ cross as a function of $\gamma$ thanks to the interactions. 

Once more, it is possible to express the ground-state energy in a general and simple form:
\begin{gather}
    \epsilon_{1}^{\text{g.s.}}=-\frac{1}{2}\left(F_{\xi}+F_{\xi}^{-1}\right)+\frac{\eta_{z}}{2\omega_{0}}
\end{gather}
with 
\begin{gather}
    F_{\xi}=\left\{\begin{array}{cc}
    f_{\xi x} & \mbox{for}\,\,\,\gamma\geq\gamma_{\xi x}^{c},
   % \,\,\, \mbox{and}\,\,\, \Delta\eta_{zx}\leq\omega_{0}, 
    \\
    f_{\xi y} & \mbox{for}\,\,\,\gamma\geq\gamma_{\xi y}^{c},
    %\,\,\, \mbox{and}\,\,\, \Delta\eta_{zy}\leq\omega_{0}, 
    \\
    1 & \mbox{otherwise} 
    \end{array}
    \right.
\end{gather}
The gradient as a function of the interactions reads
\begin{widetext}
\begin{gather}\label{eq:gradDicke}
    \nabla \epsilon_{\xi}^{\text{g.s.}}=  
    \frac{1}{2\omega_{0}}\left\{\begin{array}{cc} 
    \frac{1-f_{\xi x}^{2}}{f^{2}_{\xi x}}\left(\frac{2\omega_{0}\gamma}{\gamma_{\xi+}^{2}},-1,0,1\right)+(0,0,0,1) & \mbox{for}\,\,\,\gamma\geq\gamma_{\xi x}^{c},
    %\,\,\, \mbox{and}\,\,\, \Delta\eta_{zx}\leq\omega_{0}, 
    \\
    \frac{1-f^{2}_{\xi y}}{f^{2}_{\xi y}}\left(\frac{2\omega_{0}\gamma}{\gamma_{\xi-}^{2}},0,-1,1\right)+ (0,0,0,1)  &  \mbox{for}\,\,\, \gamma\geq\gamma_{1y}^{c},
    %\,\,\, \mbox{and}\,\,\, \Delta\eta_{zy}\leq\omega_{0}, 
    \\
   (0,0,0,1) & \mbox{otherwise} 
    \end{array}
    \right. 
\end{gather}
\end{widetext}

As a generalization of the TC and Dicke cases, we have a second-order QPT at $F_{\xi}^{c}=1$, a first-order QPTs between the superradiant-$x$ and $y$ phases at $\Delta\eta_{yz}=\Delta\eta_{zx}+\omega_{0}\left(f_{\xi +}-f_{\xi -}\right)$, and first-order QPTs from the normal to the superradiant-$x$ ($y$) phases at the values of the parameters where the other superradiant-$y$ ($x$) phase becomes prohibited. This is shown in Figs.~\ref{fig:3} (a)-(b). In Fig.~\ref{fig:3} (b) the region available for the superradiant-$x$ and $y$ phases change according to the value of $\gamma$. We are selecting the case for $\gamma/\gamma_{\xi +}=0.5$ and $\xi=0.1$ as an example. For other values of the light-matter coupling and the parameter $\xi$, we will have larger or smaller domains of validity for the superradiant phases. Similar to the two previous cases, in Figs.~\ref{fig:3} (c)-(f), we also illustrate the different behaviors and nature of the fixed points by showing the energy surfaces using the same interacting parameters as in Fig.~\ref{fig:1} and~\ref{fig:2}, but for $\xi=0.5$.

%%%%%%%%%%%%%%%%%%%%%%%%%%%%%%%%%%%%%%%%%%
%%%%%%%%%%%%%%%%%%%%%%%%%%%%%%%%%%%%%%%%%%
%%%%%%%%%%%%%%%%%%%%%%%%%%%%%%%%%%%%%%%%%%
%%%%%%%%%%%%%%%%%%%%%%%%%%%%%%%%%%%%%%%%%%

%%%%%%%%%%%%%%%%%%%%%%%%%%%%%%%%%%%%%%%%%%%%%
%%%%%%%%%%%%%%%%%%%%%%%%%%%%%%%%%%%%%%%%%%%%%
\section{Semiclassical Density of States}
\label{sec:4}
%%%%%%%%%%%%%%%%%%%%%%%%%%%%%%%%%%%%%%%%%%%%%
%%%%%%%%%%%%%%%%%%%%%%%%%%%%%%%%%%%%%%%%%%%%%

The study of quantum critical behavior is not limited to the ground-state but extends to excited energies. Next, we explore and classify the distinct energy domains emerging in each phase we have discussed in the previous section by considering the onset of Excited-State Quantum Phase Transitions (ESQPT) as a function of the light-matter and matter-matter interactions. To achieve this end, we follow the standard methodology that has been developed for the study of ESQPTs, i.e., we analyze the energy dependence and singularities of a semi-classical approximation to the Density of States (DoS) $\nu_{\xi}(\epsilon)$, obtained by calculating the available phase space volume at given energy using Weyl's law~\cite{Haake2018}
\begin{gather}\label{eq:gut}
    \nu_{\xi}(\epsilon)=\frac{1}{(2\pi)^{2}}\int \,dq\,dp\,dj_{z}\,d\phi\,\delta\left[\epsilon\omega_{0}-H_{cl}^{\xi}(q,p,j_{z},\phi)\right].
\end{gather}
Although signatures of ESQPTs can be found in the smoothed level flow, the energy densities of some observables, and the oscillatory part of the DoS~\cite{Cejnar2021,Stransky14}, the easiest way to identify them is via the smoothed DoS. 

To integrate Eq.~\ref{eq:gut}, we need to eliminate first the bosonic degrees of freedom, following closely the methodology in Refs.~\cite{Bastarrachea2014a,Bastarrachea2016}, first we clear the variable $q$ from $H_{cl}^{\xi}(q,p,j_{z},\phi)=\epsilon$ in terms of $p$, $j_{z}$ and $\phi$. Because one obtains a quadratic equation in $q$ and $p$, it always gives two roots for every value of the parameters. The two solutions are
\begin{gather}
    q_{\xi\pm}=-\frac{\gamma}{\omega}\sqrt{1-j_{z}^{2}}(1+\xi)\cos\phi
    \\ \nonumber
    \pm\sqrt{
    -p^{2}+a_{\xi}p+b_{\xi}},
\end{gather}
where
\begin{gather}
a_{\xi}=\frac{2\gamma}{\omega}\sqrt{1-j_{z}^{2}}\left(1-\xi\right)\sin\phi,\\
b_{\xi}=-\frac{2}{\omega} j_{z}(\omega_{0}+\frac{\eta_{z}j_{z}}{2})\\ \nonumber
-\frac{1}{\omega}\left(1-j_{z}^{2}\right)\left(\eta_{x}\cos^{2}\phi+\eta_{y}\sin^{2}\phi\right)\\\nonumber+\frac{2\epsilon\omega_{0}}{\omega}+\frac{\gamma^{2}}{\omega^{2}}(1-j_{z}^{2})(1+\xi)^{2}\,\cos^{2}\phi.
\end{gather}
Then, we employ the properties of the Dirac delta function to obtain 
\begin{gather}
\nu_{\xi}(\epsilon)=\frac{1}{(2\pi)^{2}}\int \,dq\,dp\,dj_{z}\,d\phi\,\\\nonumber
\left[\delta(q-q_{\xi+})\left|\frac{\partial H_{cl}^{(\xi)}}{\partial q}\right|^{-1}_{q_{\xi+}}+\delta(q-q_{\xi-})\left|\frac{\partial H^{(\xi)}_{cl}}{\partial q}\right|^{-1}_{q_{\xi-}}\right].
\end{gather}
Evaluating the derivatives leads to
\begin{gather}
\left|\frac{\partial H^{(\xi)}_{cl}}{\partial q}\right|_{q_{\xi\pm}}=\left|\omega \left[-\frac{\gamma}{\omega}\sqrt{1-j_{z}^{2}}(1+\xi)\,\cos\phi\right.\right.\\\nonumber\left.\left.\pm\sqrt{-p^{2}+a_{\xi}p+b_{\xi}}\right]+\gamma\sqrt{1-j_{z}^{2}}(1+\xi)\,\cos\phi\right|\\ \nonumber=
\omega\sqrt{-p^{2}+a_{\xi}p+b_{\xi}}.
\end{gather}
Thus, the $q$ integration yields
\begin{gather}
\nu_{\xi}(\epsilon)=\frac{1}{(2\pi)^{2}}\frac{2}{\omega}\int dj_{z}d\phi \frac{dp}{\sqrt{-p^{2}+a_{\xi}p+b_{\xi}}}.
\end{gather}
The limits in the variables $j_{z},\phi$ and $p$ are determined by the condition $-p^{2}+a_{\xi}p+b_{\xi}\geq 0$. The $p$ integration is easily performed by writing
\begin{gather}
    -p^{2}+a_{\xi}p+b_{\xi}=(p_{\xi+}-p)(p-p_{\xi-}), 
\end{gather}
where
\begin{gather}
p_{\xi\pm}=\frac{1}{2}\left(-a_{\xi}\pm\sqrt{a_{\xi}^{2}+4b_{\xi}}\right)
\end{gather}
are the roots $(p_{\xi-}\leq p_{\xi+})$ of the quadratic polynomial $-\omega^{2}p^{2}+a_{\xi}p+b_{\xi}=0$. Hence, 
\begin{gather}\label{eq:int}
    \nu_{\xi}(\epsilon)=\frac{2}{\omega(2\pi)^{2}}\int dj_{z}
    \int d\phi\\ \nonumber \int^{p_{\xi+}}_{p_{\xi-}}dp\frac{1}{\sqrt{(p_{\xi+}-p)(p-p_{\xi-})}}\\\nonumber=\frac{1}{2\pi\omega}\int dj_{z}\int d\phi
\end{gather}
This result is valid provided that the roots $p_{\xi\pm}$ are real, which, in turn, occurs only if the discriminant 
\begin{gather}
    a_{\xi}^{2}+4b_{\xi}\geq0
\end{gather}
is greater than or equal to zero. By substituting the values $a_{\xi}$ and $b_{\xi}$, this condition explicitly reads
\begin{gather}
\frac{1}{2}\left(1-j_{z}^{2}\right)\left[\left(f_{\xi+}-\frac{\eta_{x}}{\omega_{0}}\right)\cos^{2}\phi+\left(f_{\xi-}-\frac{\eta_{y}}{\omega_{0}}\right)\sin^{2}\phi\right]\nonumber\\\geq\frac{\eta_{z}}{2\omega_{0}}j_{z}^{2}+j_{z}-\epsilon,
\end{gather}
or
\begin{gather}
\cos^{2}\phi\geq g_{\xi}(j_{z},\epsilon),
\end{gather}
where
\begin{gather}
g_{\xi}(j_{z},\epsilon)=
\left\{\frac{2}{1-j_{z}^{2}}\left[\frac{\eta_{z}}{2\omega_{0}}j_{z}^{2}+j_{z}-\epsilon\right]-\left(f_{\xi-}-\frac{\eta_{y}}{\omega_{0}}\right)\right\}\nonumber\\\times\left[\left(f_{\xi+}-f_{\xi-}\right)-\left(\frac{\eta_{x}}{\omega_{0}}-\frac{\eta_{y}}{\omega_{0}}\right)\right]^{-1}.
\end{gather}
We observe these expressions only depend on the phase space volume over the region of the Bloch sphere covered at a given energy. They allow us to determine the limiting values for $(j_{z},\phi)$ in the Bloch sphere, given that $0\leq\cos\phi_{0}\leq1$. If $f_{\xi}(j_{z},\epsilon)<0$, then the condition can be satisfied for all the values of $\phi\in[0,2\pi)$, covering the Bloch sphere. Instead, if $f_{\xi}(j_{z},\epsilon)>1$ the condition cannot be fulfilled. It would be valid only within an interval of $\phi$ given by the limiting angle 
\begin{gather}\label{eq:xidos1}
\phi_{\xi}=\arccos\sqrt{g_{\xi}(j_{z},\epsilon)}\\ \nonumber
=\arccos\left\{\left[\frac{2}{1-j_{z}^{2}}\left[\frac{\eta_{z}}{2\omega_{0}}j_{z}^{2}+j_{z}-\epsilon\right]-\left(f_{\xi-}-\frac{\eta_{y}}{\omega_{0}}\right)\right]^{1/2}\right.\\\nonumber\times\left.
\left[\left(f_{\xi+}-f_{\xi-}\right)-\left(\frac{\eta_{x}}{\omega_{0}}-\frac{\eta_{y}}{\omega_{0}}\right)\right]^{-1/2}
\right\}.
\end{gather}
such that $[-\phi_{\xi},\phi_{\xi}]$ or $[\pi-\phi_{\xi},\pi+\phi_{\xi}]$. 
In general, we can obtain limiting values for $j_{z}$ and $\varepsilon$ where the condition is satisfied by taking into account the aforementioned limits of $\cos{\phi}$. 

First, we consider the limits given by $\cos\phi_{\pm}=\pm 1$. It leads to a quadratic equation for $j_{z}$ which reads
\begin{gather}\label{eq:jzq}
j_{z}^{2}\left\{\frac{\eta_{z}}{2\omega_{0}}+\frac{1}{2}\left(f_{\xi+}-\frac{\eta_{x}}{\omega_{0}}\right)\right\}+j_{z}\\\nonumber-
\left\{\epsilon+\frac{\eta_{z}}{2\omega_{0}}+\frac{1}{2}\left(f_{\xi+}-\frac{\eta_{z}-\eta_{x}}{\omega_{0}}\right)\right\}
=0,
\end{gather}
where we have inserted a zero by adding and substracting $\eta_{z}/2\omega_{0}$. We observe that the effect of the interactions in $z$ direction is to shift the energy, while in the $x$ direction is to shift the critical coupling. The resulting roots are
\begin{gather} \label{eq:xidos2}
   j_{z\xi}^{(\pm)}(\varepsilon)=-\frac{1}{f_{\xi x}}\left[1\mp\sqrt{2f_{\xi x}(\epsilon-\epsilon_{s\xi x})}\right]
\end{gather}
Second, we obtain the limits given by $\cos\phi_{1,2}=0$. Likewise, we get a quadratic equation which reads
\begin{gather}
j_{z}^{2}\left[\frac{\eta_{z}}{2\omega_{0}}+\frac{1}{2}\left(f_{\xi-}-\frac{\eta_{y}}{\omega_{0}}\right)\right]+j_{z}\\\nonumber-\left[\epsilon+\frac{\eta_{z}}{2\omega_{0}}+\frac{1}{2}\left(f_{\xi-}-\frac{\eta_{z}-\eta_{y}}{\omega_{0}}\right)\right]=0.
%\left[\frac{\eta_{z}}{2}j_{z}^{2}+j_{z}-\epsilon\right]-\frac{1-j_{z}^{2}}{2}\left(f_{\xi-}-\frac{\eta_{y}}{\omega_{0}}\right)=0.
\end{gather}
where we have also added and subtracted $\eta_{z}/2\omega_{0}$. We notice it is identical to Eq.~\ref{eq:jzq}, but changing $\eta_{x}\rightarrow\eta_{y}$ and $f_{\xi -}\rightarrow f_{\xi +}$. Consequently, the solutions are given by
\begin{gather} \label{eq:xidos3}
   j_{z\xi}^{(1,2)}(\varepsilon)=-\frac{1}{f_{\xi y}}\left[1\mp\sqrt{2f_{\xi y}(\epsilon-\epsilon_{s\xi y})}\right].
\end{gather}
In the following, we will study the particular cases of the TC ($\xi=0$) and Dicke ($\xi=1$) to understand the effects of the interactions on the emergence of energy domains and critical energies. Finally, we will comment about the arbitrary $\xi$ case and the general typology of ESQPTs. 

%%%%%%%%%%%%%%%%%%%%%%%%%%%%%%%%%%%%%%%%%%%%%
\subsection{Energy domains in the TC limit}
%%%%%%%%%%%%%%%%%%%%%%%%%%%%%%%%%%%%%%%%%%%%%

In the TC limit, the key functions determining the integral in Eq.~\ref{eq:int} are given by
\begin{gather} \label{eq:tccdos}
\phi_{0}(j_{z},\varepsilon)=\arccos\left\{
\left[\frac{2}{1-j_{z}^{2}}\left(\frac{\eta_{z}}{2\omega_{0}}j_{z}^{2}+j_{z}-\varepsilon\right)\right.\right.\\\nonumber\left.\left.-\left(f_{0+}-\frac{\eta_{y}}{\omega_{0}}\right)\right]^{1/2}\left(\frac{\eta_{y}}{\omega_{0}}-\frac{\eta_{x}}{\omega_{0}}\right)^{-1/2}
\right\},\\
   j_{z0}^{(\pm)}(\varepsilon)=-\frac{1}{f_{0x}}\left[1\mp\sqrt{2f_{0x}(\epsilon-\epsilon_{s0x})}\right],\\
   j_{z0}^{(1,2)}(\varepsilon)=-\frac{1}{f_{0y}}\left[1\mp\sqrt{2f_{0y}(\epsilon-\epsilon_{s0y})}\right].
\end{gather}

We observe the expressions for the critical coupling and the ground-state energy of the superradiant-$x$ and $y$ phases of the TC limit are recovered when one considers either $j_{z0}^{(\pm)}$ or $j_{z0}^{(1,2)}$, respectively. We note that $j_{z0}^{(1)}\leq j_{z0}^{(2)}$. We must compare these values with those coming from the stationary points $j_{z}=\pm 1$, i.e., the ones that are present in all the phases and whose energy is given by $\varepsilon_{\pm}=\pm 1+\eta_{z}/2\omega_{0}$. Subsequently, we recognize four different energy phases and three critical energies using the comparison between the values of $j_{z}$, the conditions in Eq.~\ref{eq:tccdos}, and what we learned in Sec.~\ref{sec:3}. These energy domains correspond to various behaviors of the function $g_{\xi}(j_{z},\epsilon)$. In turn, they determine the intervals of the variable $j_{z}$. Without loss of generality, let's assume that we select $\eta_{x}$ and $\eta_{y}$ such that $\epsilon_{s0x}<\epsilon_{s0y}$. Then, the energy phases are:
\begin{enumerate}
\item The upper interval where $\epsilon_{+}<\epsilon$. Here, the function $g_{0}(j_{z},\epsilon)$ is always lesser than one. The whole pseudospin sphere is available: $j_{z}\in[-1,1]$ and $\phi_{0}\in[0,2\pi)$. Consequently, the available phase space volume (per $j$) saturates to its limiting value $\nu_{0}(\epsilon)=2/\omega$.
\item The interval where $\epsilon_{-}<\epsilon<\epsilon_{+}$. Here, $j_{z}$  takes values only in the interval $\left[-1,j_{z0}^{(+)}\right]$ with $\left|j_{z0}^{(+)}\right|\leq1$. This interval is always present for all values of parameters and corresponds to available phase space from the absolute minimum point at $j_{z}=-1$ and the absolute maximum at $j_{z}=+1$. %
\item The interval that is only present in the superradiant-$y$ phase, $\epsilon_{0sy}<\epsilon\leq\epsilon_{-}$. 
\item The interval arising in presence of both the superradiant-$x$ and $y$ phases, $\epsilon_{0sx}\leq\epsilon\leq\epsilon_{0sy}$. Here, the south pole of the pseudospin sphere ($j_{z}=-1$) is inaccessible and the variable $j_{z}$ is restricted to the interval $j_{z0}^{(-)}\leq 1 \leq j_{z0}^{(+)}$. Considering that $\epsilon_{0sx}<\epsilon_{0sy}$ is the ground-state energy in the superradiant-$x$ phase.
\end{enumerate}
Clearly, we have three critical energies given by $\epsilon_{0}^{(c1)}=\epsilon_{s0y}$, $\epsilon_{0}^{(c2)}=\epsilon_{-}$, and $\epsilon_{0}^{(c1)}=\epsilon_{+}$. All of them correspond to stationary points of the energy surface and to what we have already found in Sec.~\ref{sec:3}. The semiclassical approximation to the DoS in the TC model becomes
\begin{widetext}
\begin{gather}
\frac{\omega}{2}\nu_{0}(\epsilon)=
    \left\{\begin{array}{lc}
       \frac{1}{\pi}\int_{j_{z0}^{(-)}}^{j_{0}^{(+)}}\phi_{0}(j_{z},\epsilon)dj_{z}, & \epsilon\in[\epsilon_{0sx},\epsilon_{0sy}]\,\,\,\mbox{and}\,\,\,\gamma\in[\gamma_{0x}^{c},\gamma_{0y}^{c}], \\
         \frac{1}{\pi}\left[\int_{j_{z0}^{(-)}}^{j_{z0}^{(1)}}\phi_{0}(j_{z},\epsilon)dj_{z}+\int_{j_{z0}^{(2)}}^{j_{z0}^{(+)}}\phi_{0}(j_{z},\epsilon)dj_{z}\right]+\frac{1}{2}\left(j_{z0}^{(2)}-j_{z0}^{(1)}\right), & \epsilon\in[\epsilon_{s0y},\epsilon_{-}]\,\,\,\mbox{and}\,\,\,\gamma\in[\gamma_{0y}^{c},\infty),\\
         \frac{1}{\pi}\int_{j_{z0}^{(1)}}^{j_{z0}^{+}}\phi_{0}(j_{z},\epsilon)dj_{z}+\frac{1}{2}\left(j_{z0}^{(1)}+1\right), & \epsilon\in[\epsilon_{-},\epsilon_{+}], \mbox{and}\,\,\,\gamma\in[0,\infty),\\
         1, & \epsilon_{+}\leq\epsilon,\mbox{and}\,\,\,\gamma\in[0,\infty).
    \end{array}\right.
\end{gather}
\end{widetext}
The onset of these energy domains depends on the three intervals of $\gamma$ that we discussed in Sec~\ref{sec:3}. 
The boundary between each energy domain signals the existence of an ESQPT, as the DoS has a critical change characterized by a singularity in its derivative, even though the DoS is continuous in the energy variable. The type of ESQPT is encoded in the first derivative of the DoS, 
$d\nu_{0}(\varepsilon)/d\epsilon$,
which in turn is in terms of the derivative 
\begin{gather}
\frac{\partial\phi_{0}}{\partial\epsilon}=\frac{1}{1-j_{z}^{2}}\left\{\left[1-g_{0}(j_{z},\varepsilon)\right]g_{0}\left(j_{z},\varepsilon\right)\left(\frac{\eta_{y}-\eta_{x}}{\omega_{0}}\right)\right\}^{-1/2}.
\end{gather}
It can be shown that those ESQPTs at $\epsilon_{0sy}$ corresponds to a logarithmic-type discontinuity, and the ones at $\epsilon_{-}$ and $\epsilon_{+}$ are of the jump-type. This is not the typical behavior of the standard TC model: we recover it only when $\eta_{x}=\eta_{y}$. There, the symmetry leads to two jump-type singularities at $\epsilon_{0sy}=\epsilon_{0sx}$ and $\epsilon_{+}$~\cite{Brandes13,Bastarrachea2014a}. This is because $\epsilon_{0sy}$ becomes the ground-state, so only the ESQPTs corresponding to the fixed points at $j_{zs}=\pm1$ remain. We will offer a unified explanation of this behavior later, when we discuss the arbitrary $\xi$ case. The volume of the available phase space for the TC model, encoded in the form of the semiclassical DoS, is shown for three different sets of interacting parameters, as a function of the energy, in the top row of Fig.~\ref{fig:4} (a)-(c), where we have chosen interacting parameters to highlight the different domains. 

%%%%%%%%%%%%%%%%%%%%%%%%%%%%%%%%%%%%%%%%%%%%%
\subsection{Energy domains in the Dicke limit}
%%%%%%%%%%%%%%%%%%%%%%%%%%%%%%%%%%%%%%%%%%%%%

Following the same reasoning as in the previous section, we derive the expressions for $\xi=1$:
\begin{gather} \label{eq:dicdos}
\phi_{1}(j_{z},\varepsilon)=
\arccos\left\{\left[\frac{2}{1-j_{z}^{2}}\left[\frac{\eta_{z}}{2\omega_{0}}j_{z}^{2}+j_{z}-\epsilon\right]+\frac{\eta_{y}}{\omega_{0}}\right]^{1/2}\right. \nonumber \\ \left.\times
\left(f_{1+}-\left(\frac{\eta_{x}}{\omega_{0}}-\frac{\eta_{y}}{\omega_{0}}\right)\right]^{-1/2}
\right\},\\
   j_{z1}^{(\pm)}(\varepsilon)=-\frac{1}{f_{1x}}\left[1\mp\sqrt{2f_{1x}(\epsilon-\epsilon_{s1x})}\right],\\
   j_{z1}^{(1,2)}(\varepsilon)=-\frac{1}{f_{1y}}\left[1\mp\sqrt{2f_{1y}(\epsilon-\epsilon_{s1y})}\right].
\end{gather}
where $j_{z1}^{(1)}\leq j_{z1}^{(2)}$. We recall that $f_{1y}$ is independent of the light-matter coupling. For the Dicke model, the range of the $j_{z}$ variable is given by the same expressions as in the TC model. Thus, we get the following intervals:
\begin{enumerate}
\item The interval $\epsilon_{+}<\epsilon$, where, as in the TC model the whole pseudospin sphere is available $j_{z}\in[-1,1]$, $\phi_{1}\in[0,2\pi)$, and $\nu_{1}(\epsilon)=2/\omega$.
\item The interval $\epsilon_{-}<\epsilon<\epsilon_{+}$. Here the $j_{z}$ variable takes values only in the interval $[-1,j_{z1}^{(+)}]$ and $\phi_{1}$ is restricted. When $j_{z}\in[-1,\epsilon]$, $\phi_{1}$ takes values in the whole interval $[0,2\pi)$, but if $\epsilon<j_{z}\leq j_{z1}^{(+)}$, $0<\phi_{1}<\pi$.
\item The interval $\epsilon_{1sy}<\epsilon\leq\epsilon_{-}$. It only appears in the deformed phase. 
\item The lower interval  $\epsilon_{1sx}\leq\epsilon\leq\epsilon_{1sy}$. Here, the south pole of the Bloch sphere $(j_{zs}=-1)$ is inaccessible and the $j_{z}$ variable becomes restricted to the interval $j_{z}\in[j_{z1}^{(-)},j_{z1}^{(+)}]$.
\end{enumerate}
The expression for the semiclassical DoS in the Dicke limit becomes
\begin{widetext}
\begin{gather}
\frac{\omega}{2}\nu_{1}(\epsilon)=
    \left\{\begin{array}{lc}
         \frac{1}{\pi}\int_{j_{z1}^{(-)}}^{j_{z1}^{(+)}}\phi_{1}(j_{z},\epsilon)dj_{z}, & \epsilon\in[\epsilon_{s1x},\epsilon_{s1y}]\,\,\,\mbox{and}\,\,\,\gamma\in[\gamma_{1x}^{c},\infty), \\
         \frac{1}{\pi}\left[\int_{j_{z1}^{(-)}}^{j_{z1}^{(1)}}\phi_{1}(j_{z},\epsilon)dj_{z}+\int_{j_{z1}^{(2)}}^{j_{z1}^{(+)}}\phi_{1}(j_{z},\epsilon)dj_{z}\right]+\frac{1}{2}\left(j_{z1}^{(2)}-j_{z1}^{(1)}\right), & \epsilon\in[\epsilon_{s1y},\epsilon_{-}] \,\,\,\mbox{and}\,\,\,\gamma\in[\gamma_{1x}^{c},\infty],\\
         \frac{1}{\pi}\int_{j_{z1}^{(1)}}^{j_{z1}^{(+)}}\phi_{1}(j_{z},\epsilon)dj_{z}+\frac{1}{2}\left(j_{z1}^{(1)}+1\right), & \epsilon\in[\epsilon_{-},\epsilon_{+}]\,\,\,\mbox{and}\,\,\,\gamma\in[0,\infty],\\
         1, & \epsilon_{+}\leq\epsilon\,\,\,\mbox{and}\,\,\,\gamma\in[0,\infty].
    \end{array}\right.
\end{gather}
\end{widetext}
If we cancel the interactions in $z$ and $y$ directions ($\eta_{x}=\eta_{y}$), we recover Eq. 19 in Ref.~\cite{Rodriguez2018}, where $\epsilon_{s1y}=-\omega_{0}/2\eta_{z}$ and
\begin{gather} 
\phi_{1}(j_{z},\varepsilon)=
\arccos\left\{\left[\frac{2}{1-j_{z}^{2}}\left(\frac{\eta_{z}}{2\omega_{0}}j_{z}^{2}+j_{z}-\epsilon\right)\right]^{1/2}
f_{1+}^{-1/2}
\right\}.
\end{gather}

The volume of the available phase space for Dicke model for three different couplings, as a function of the energy, is shown in the middle row of Fig.~\ref{fig:4} (d)-(f) for representative values of the parameters. The singular behavior of the DoS is encoded in the derivative
\begin{gather}
\frac{\partial\phi_{1}}{\partial\epsilon}=\frac{1}{1-j_{z}^{2}}\left\{\left[1-g_{1}(j_{z},\varepsilon)\right]g_{1}(j_{z},\varepsilon)\right.\\\nonumber\left.\times\left(f_{1+}-\frac{\eta_{x}-\eta_{y}}{\omega_{0}}\right)\right\}^{-1/2}.
\end{gather}

Even though the fixed points belonging to the deformed phase do not appear as extrema in the superradiant-$x$ phase, they still impact the energy domains, given that they are related to the interactions in the $z$ directions via $j_{z1}^{(1,2)}$. Then, one can still find four different energy domains and three ESPQT. As it is shown in Fig.~\ref{fig:4} (middle row), the DoS curves for the Dicke and TC limits are very similar as a function of the energy $\epsilon$ for small couplings. The behavior becomes analogous to the TC case. Although this result is similar to that of an extended Dicke model, it differs from the standard Dicke model, where the singularity at $\epsilon_{-}$ is of the logarithmic type~\cite{Brandes13,Bastarrachea2014a}.

%%%%%%%%%%%%%%%%%%%%%%%%%%%%%%%%%%%%%%%%%%%%%
\subsection{Energy domains for arbitrary couplings and typology of ESQPTs}
%%%%%%%%%%%%%%%%%%%%%%%%%%%%%%%%%%%%%%%%%%%%%

Finally, we offer some considerations about the most general case. When an arbitrary value of $\xi\in(0,1)$ is chosen, we obtain a general expression by combining Eqs.~\ref{eq:xidos1},~\ref{eq:xidos2}, and~\ref{eq:xidos3}. It reads
\begin{widetext}
\begin{gather}
\frac{\omega}{2}\nu_{\xi}(\epsilon)=
    \left\{\begin{array}{lc}
         \frac{1}{\pi}\int_{j_{z\xi}^{(-)}}^{j_{z\xi}^{(+)}}\phi_{\xi}(j_{z},\epsilon)dj_{z}, & \epsilon_{s\xi x}\leq\epsilon\leq\epsilon_{s\xi y},\,\,\,\mbox{and}\,\,\,\gamma\in[\gamma_{\xi x}^{c},\gamma_{\xi y}^{c}], \\
         \frac{1}{\pi}\left[\int_{j_{z\xi}^{(-)}}^{j_{z}^{(1)}}\phi_{0}(j_{z},\epsilon)dj_{z}+\int_{j_{z\xi}^{(2)}}^{j_{z\xi}^{(+)}}\phi_{0}(j_{z},\epsilon)dj_{z}\right]+\frac{1}{2}\left(j_{z\xi}^{(2)}-j_{z\xi}^{(1)}\right), & \epsilon_{s\xi y}<\epsilon\leq\epsilon_{-},\,\,\,\mbox{and}\,\,\,\gamma\in[\gamma_{\xi y}^{c},\infty], \\
         \frac{1}{\pi}\int_{j_{z\xi}^{(1)}}^{j_{z\xi}^{(+)}}\phi_{0}(j_{z},\epsilon)dj_{z}+\frac{1}{2}\left(j_{z\xi}^{(1)}+1\right), & \epsilon_{-}<\epsilon\leq\epsilon_{+},\,\,\,\mbox{and}\,\,\,\gamma\in[0,\infty), \\
         1, & \epsilon_{+}<\epsilon\,\,\,\mbox{and}\,\,\,\gamma\in[0,\infty). 
    \end{array}\right.
\end{gather}
\end{widetext}
Given that $\epsilon_{s\xi x}<\epsilon_{s\xi y}$. This expression reunites the effects of the qubit-qubit interactions and the arbitrary light-matter coupling $\xi$. As discussed before, the phenomenology of this result is similar to what is found in absence of interactions but for arbitrary $\xi$ as shown in Refs.~\cite{Bastarrachea2016,Kloc2017,Stransky2017b,Cejnar2021}. The main difference being the modification of the critical coupling of the superradiant-$(+)$ phase by $\eta_{x}$, the critical coupling of the superradiant-$(-)$ phase by $\eta_{y}$ and the direct shift of the energy and $j_{z}$ intervals of validity by $\eta_{z}$. Also, in this case we have as a general expression 
\begin{gather} 
\frac{\partial\phi_{\xi}}{\partial\epsilon}=\frac{1}{1-j_{z}^{2}}\left\{\left[1-g_{\xi}(j_{z},\varepsilon)\right]g_{\xi}\left(j_{z},\varepsilon\right)\right.\\\nonumber\left.\left[\left(f_{\xi+}-f_{\xi-}\right)-\frac{\eta_{x}-\eta_{y}}{\omega_{0}}\right]\right\}^{-1/2}.
\end{gather}

ESQPTs can be classified using of two numbers: the index of the transition $r$, which denotes the number of negative eigenvalues of the Hessian matrix of the classical Hamiltonian at the stationary points (where the phase space volume change), and the number of relevant degrees of freedom $f$ of the system, determining in which derivative of the smooth DoS the discontinuity associated to the ESQPT appears~\cite{Stransky2016}. This classification is tied to the properties of the Hessian matrix describing the local dependence of the Hamiltonian around the fixed points of the energy surface, so it is valid only if the Hessian does not have zero or singular eigenvalues. $r$ determines the type of singularity: for even $f$ systems, $r=1$ a logarithmic-type singularity, while $r=2$ means a jump-type one~\cite{Stransky2016,Stransky2017b,Cejnar2021}. Because the Dicke model has only two degrees of freedom (the collective spin and the boson), it has $f=2$. The integrability of the standard TC model reduces to $f=1$ instead~\cite{Stransky2017b}. For arbitrary $\xi$ it has been shown that there are three ESPQT marked by three critical energies given by $\epsilon_{\xi}^{c1}=\epsilon_{s\xi+}$, $\epsilon_{\xi}^{c2}=\epsilon_{-}$, and $\epsilon_{\xi}^{c3}=\epsilon_{+}$. Their indices correspond $r=1$, $r=2$, and $r=2$, respectively, corresponding to saddle points ($r=1$) or maxima $(r=2)$, as discussed in Sec.~\ref{sec:3}. 

\begin{widetext}
\begin{figure*}
\begin{center}
\begin{tabular}{c}
\includegraphics[width=0.8\textwidth]{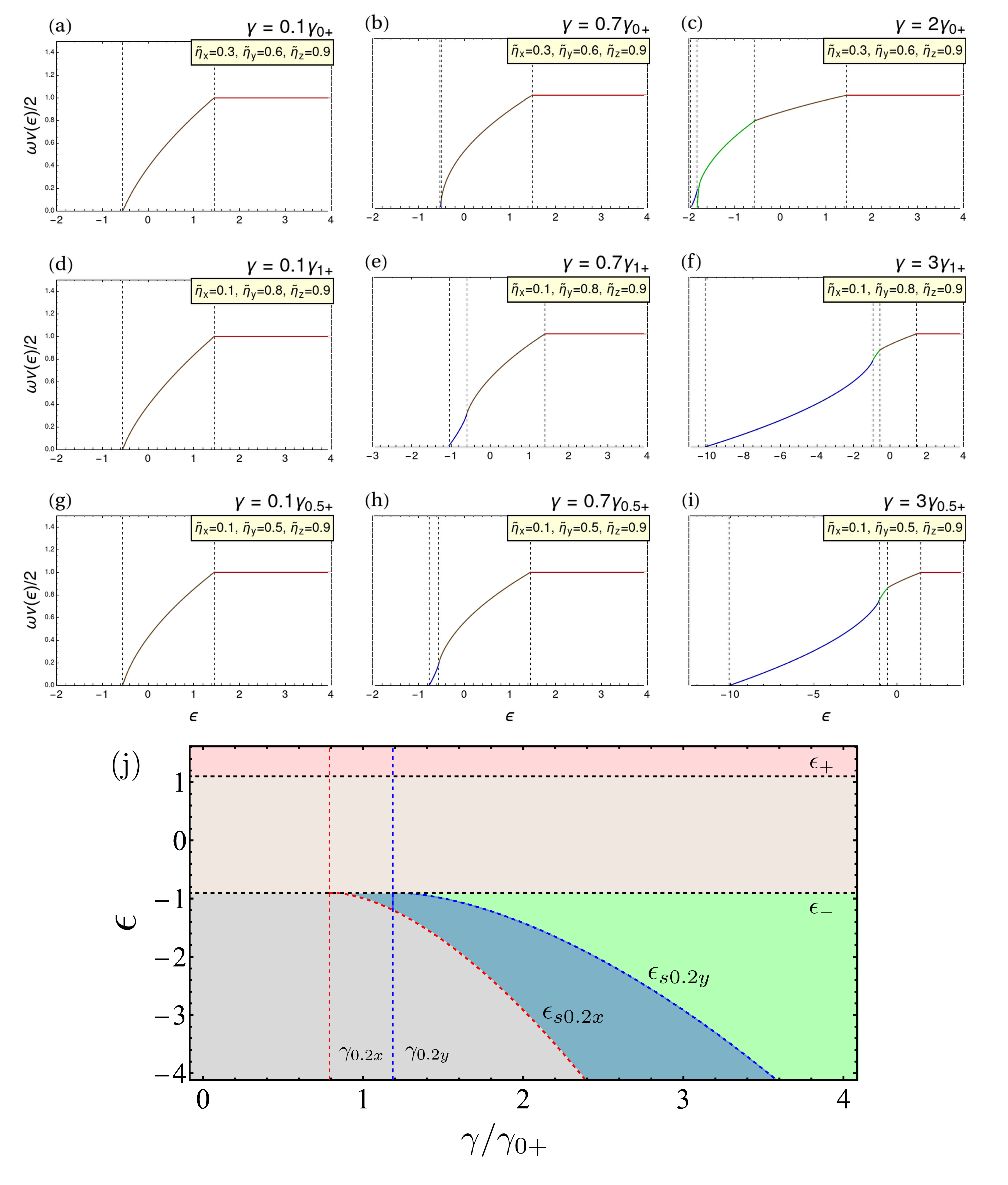}
\end{tabular}
\end{center}
\vspace{-20pt}
\caption{(a)-(i) Density of States $\omega\nu_{\xi}(\epsilon)/2$ as a function of energy for the TC limit (top row), the Dicke limit (middle row), and an arbitrary coupling set at $\xi=0.5$ (bottom row), for several values of the light-matter coupling and the qubit-qubit interactions chosen to highlight the different energy domains. We assume the case where the superradiant-$x$ phase is below the superradiant-$y$ phase. Thus, we exhibit three general regimes: normal (left column), superradiant-$x$ (middle column), and superradiant-$x$ modified by the fixed points from the superradiant-$y$ or deformed phases (right column). The four energy domains that can be encountered are marked with different colors: $[\epsilon_{s\xi x},\epsilon_{s\xi y}]$ (blue), $[\epsilon_{s\xi y},\epsilon_{-}]$ (green), $[\epsilon_{-},\epsilon_{+}]$ (brown), $[\epsilon_{+},\infty)$ (red). The relevant energies, including the ground-state and critical ones, are indicated with black vertical dashed lines. (j) Diagram of the energy domains as a function of $\gamma$ for $\xi=0.2$, $\eta_{x}=\eta_{y}=1$ and $\eta_{z}=2$. The colors indicate the corresponding domain to Figs.~\ref{fig:4} (a)-(i). The relevant energies are indicated with dashed curves and the critical couplings with dashed vertical lines.}
\label{fig:4}
\end{figure*}
\end{widetext}

A major result of our exploration is that the interactions in $\eta_{x}$ and $\eta_{y}$ play a similar role to an arbitrary value of $\xi$ in both the ground-state and excited-state properties. $\eta_{x}$ modifies the magnitude of $\gamma_{\xi+}^{c}$, and $\eta_{y}$ does the corresponding for $\gamma_{\xi-}^{c}$. As a result, the ESQPT coming from the transition between the superradiant-$x$ and superradiant-$y$ domains (and vice versa) are analogous to the transition between the superradiant-$(+)$ and $(-)$ phase at $\eta_{i}=0$. This ESQPT has an index $r=1$~\cite{Kloc2017}. Likewise, for an arbitrary value of matter-matter interactions and $\xi$ both ESQPTs, the one at the stationary point $j_{zs}=-1$ and the one due to the saturation of the Bloch sphere ($j_{zs}=+1$) have $r=2$~\cite{Bastarrachea2014a,Kloc2017,Stransky2017b}. This explains our findings for the TC and Dicke limits. In the first case, the extended TC model, including interactions, is not integrable anymore. In the Dicke case, we could still have a finite $\eta_{x}$ modifying the DoS even though the critical coupling $\gamma_{1-}^{c}$ goes to infinity. Therefore, as it is revealed in Fig.~\ref{fig:4} (g)-(i), where we use $\xi=0.5$ and various values of the qubit interactions as representative examples, we will find a similar typology to the cases we have mentioned in the absence of qubit-qubit interactions. Finally, in Fig.~\ref{fig:4} (j) we illustrate the possible energy domains and the location of the critical energies as a function of $\gamma$ for  $\xi=0.2$, $\eta_{x}=\eta_{y}=1$ and $\eta_{z}=2$.  

%%%%%%%%%%%%%%%%%%%%%%%%%%%%%%%%%%%%%%%%%%%%%
%%%%%%%%%%%%%%%%%%%%%%%%%%%%%%%%%%%%%%%%%%%%%
\section{Discussion and conclusions}
\label{sec:5}
%%%%%%%%%%%%%%%%%%%%%%%%%%%%%%%%%%%%%%%%%%
%%%%%%%%%%%%%%%%%%%%%%%%%%%%%%%%%%%%%%%%%%

In this work, we have investigated the quantum phases emerging in an extended Dicke model that involves qubit-qubit interactions. We have also included the possibility of varying the strength of the non-resonant terms so that the system can go from the Tavis-Cummings to the Dicke regimes. To achieve this end, we have used standard semiclassical techniques, whose central element is considering the expectation value of the quantum generalized Hamiltonian over a tensor product of Bloch and Glauber coherent states. By studying the shape of the energy surfaces, their stationary points, and the behavior of the semiclassical approximation to the Density of States, one can identify and characterize the QPT, ESQPTs, quantum phases, and energy domains resulting from the combination of light-matter (spin-boson) and matter-matter (collective spin) interactions. 

We have found general expressions for the ground-state energy and analyzed the QPTs as a function of the Hamiltonian parameters in three cases: for the Tavis-Cummings limit ($\xi=0$), the Dicke limit ($\xi=1$), and for an arbitrary interaction strength in between these two. We have considered a general combination of collective qubit interactions represented by operators $\hat{J}_{i}^{2}$ with strengths $\eta_{i}$ and $i=x,y,z$. This is the most general case for two-body interactions between the collective degrees of freedom of the qubits. Each direction has a particular role in modifying the critical phenomena of the standard light-matter system for both the ground and excited states. To start, we examine the results for interactions in the $z$ direction. As mentioned before, three main results have been discovered before due to a finite $\eta_{z}$: shifting of the ground-state by $\eta_{z}/2$, the onset of first-order phase transitions, and the modification of the critical value of the light-matter interaction where the superradiant QPT appears (see, e.g., Ref.~\cite{Rodriguez2018}). Indeed, we have confirmed these results and generalized them as we have found that the same phenomena occur in the presence of interactions in the $x$ and $y$ directions. Also, we have noted that the relevant parameters of the system are the differences $\Delta\eta_{z,x}$ and $\Delta\eta_{z,y}$, which is a natural result due to the conservation of the pseudospin length. Tuning these quantities allows for the stimulation and suppression of superradiance via manipulating the light-matter interaction.

However, this is not the only effect of the $x$ and $y$ interactions. In terms of the $\Delta\eta_{zi}$ ($i=x,y$) parameters, they produce two new quantum phases, the superradiant-$x$ and superradiant-$y$ phases. If we assume the interactions in the $x$ and $y$ directions are balanced, we recover the distinctive rotational symmetry of the standard TC model. No matter the value of $\xi$, the normal phase would be symmetric, and, in the case of the TC limit, the superradiant phase will correspond to the well-known Mexican hat potential. Thus, we call it the superradiant-symmetric phase. In the imbalanced case, we observe new effects. The integrability of the TC Hamiltonian breaks down, and the two superradiant phases appear. Besides, the energy surface of the normal phase is deformed for every $\xi$ stretching in the $x$ or $y$ direction depending on the relationship between $\Delta\eta_{zx}$ and $\Delta\eta_{zy}$. Additionally, new effects appear, such as first-order QPT between the $x$, $y$, and normal phases and the existence of parameter domains where the fixed points of both the $x$ and $y$ phases coexist. We refer to this situation as a superposition of phases, where one of them can be dominant~\cite{Bastarrachea2016}. The passage between a single superradiant phase to one in a superposition does not imply a QPT because the ground-state remains the same. Still, it will affect the energy domains and ESQPT present for that specific parameter set. On the other hand, the Dicke limit becomes a situation where the superradiant-$y$ phase vanishes and leaves a deformed or subradiant phase first identified in Ref.~\cite{Rodriguez2018}. It only occurs for $\Delta\eta_{zy}\geq1$ independently of $\gamma$. The onset of this phase produces the development of a new first-order QPT between it, the superradiant-$x$, and normal phases. Also, it suppresses any superposition between phases.

Notoriously, one can understand these results from a unified point of view by looking at the arbitrary $\xi$ case in general. In the absence of interactions, an intermediate value of the light-matter interaction leads to the existence of two phases, the superradiant-$(+)$ and $(-)$. Their position in the quantum phases landscape is fixed, depending on the relationship between the critical couplings $\gamma_{\xi\pm}^{c}$. As a result, for a light-matter interaction larger than $\gamma_{\xi-}^{c}$, the two phases are superimposed~\cite{Bastarrachea2016,Kloc2017,Stransky2017b,Cejnar2021}. It turns out the superradiant-$x$ ($y$) phase is a generalization of the superradiant-$(+)$ [$(-)$] phase. Therefore, the phenomenology of critical phenomena for both the ground and excited states is similar. This has been confirmed by analyzing the semiclassical Density of States in the three regimes of the light-matter coupling and for the various cases of qubit interaction strengths. We have obtained general expressions for the DoS and the limiting values of the atomic variables $(j_{z},\phi)$ in the Bloch sphere that let to identify energy domains and critical energies tied to ESQPTs. Finally, we have unveiled a unique feature due to the qubit interactions. Unlike the superradiant-$(\pm)$, the landscape of the superradiant $x$ and $y$ phases can be modified at will by independently tuning the qubit interaction strengths. This specific feature is left to be studied in the near future.

Our study provides a broad perspective of critical phenomena in collective models combining strong light-matter and matter-matter interactions. Future directions like the exploration of the existence and robustness of Goldstone and Higgs modes in quantum optical setups~\cite{YiXiang2013} may benefit from the general description of the quantum phases our results provide. Moreover, as experimental progress promises to make individually controlled interactions in each direction feasible soon, we expect our work to be a reference for exploring critical quantum phenomena in quantum information, atomic physics, quantum optics, and condensed matter systems involving collective qubits interactions. 

%%%%%%%%%%%%%%%%%%%%%%%%%%%%%%%%%%%%%%%%%%
\vspace{6pt} 
\acknowledgments{M.A.B.M. acknowledges fruitful discussions with J. G. Hirsch and S. A. Lerma-Hern\'andez in the early stages of this project.%, and with N. Ram\'irez-Cruz
}
%%%%%%%%%%%%%%%%%%%%%%%%%%%%%%%%%%%%%%%%%%

%%%%%%%%%%%%%%%%%%%%%%%%%%%%%%%%%%%%%%%%%%
%\conflictsofinterest{The authors declare no conflict of interest. The funders had no role in the design of the study; in the collection, analyses, or interpretation of data; in the writing of the manuscript, or in the decision to publish the results.} 

%%%%%%%%%%%%%%%%%%%%%%%%%%%%%%%%%%%%%%%%%%
%% optional
%\abbreviations{The following abbreviations are used in this manuscript:\\

%\noindent 
%\begin{tabular}{@{}ll}
%DoS & Density of States \\
%QPT & Quantum Phase Transition \\
%ESQPT & Excited-State Quantum Phase Transition \\
%TC & Tavis-Cummings \\
%LMG & Lipking-Meshkov-Glick \\
%SC & Strong coupling \\
%USC & Ultra-strong coupling 
%\end{tabular}}

%%%%%%%%%%%%%%%%%%%%%%%%%%%%%%%%%%%%%%%%%%
%% optional
%\appendixtitles{yes} %Leave argument "no" if all appendix headings stay EMPTY (then no dot is printed after "Appendix A"). If the appendix sections contain a heading then change the argument to "yes".
\appendix
\unskip

%%%%%%%%%%%%%%%%%%%%%%%%%%%%%%%%%%%%%%%%%%
%%%%%%%%%%%%%%%%%%%%%%%%%%%%%%%%%%%%%%%%%%
\section{Hamilton Equations for the TC and Dicke limits}
\label{app:a}
%%%%%%%%%%%%%%%%%%%%%%%%%%%%%%%%%%%%%%%%%%
%%%%%%%%%%%%%%%%%%%%%%%%%%%%%%%%%%%%%%%%%%

Here, we present Hamilton equations Eqs.~\ref{eq:he1} to~\ref{eq:he4} for $\xi=0$. They turn out to be:
\begin{gather}
    \dot{q}=\omega p-\gamma\sqrt{1-j^{2}_{z}}\sin\phi,\,\,\,\\ 
    \dot{p}-\omega q-\gamma\sqrt{1-j^{2}_{z}}\cos\phi,\\
   \dot{\phi}=\omega_{0}+\eta_{z}j_{z}-j_{z}(\eta_{x}\cos^{2}\phi+\eta_{y}\sin^{2}\phi)
   \\ \nonumber
    -\frac{\gamma j_{z}}{\sqrt{1-j^{2}_{z}}}\left(q\cos\phi-p\sin\phi\right),\\
    \dot{j_{z}}=\left(1-j^{2}_{z}\right)(\eta_{x}-\eta_{y})\cos\phi\sin\phi+
    \\ \nonumber
    \gamma\sqrt{1-j^{2}_{z}}\left(q\sin\phi+p\cos\phi\right).
\end{gather}
Likewise, the Hamilton Equations for $\xi=1$ are
\begin{gather}
    \dot{q}=\omega p, \,\,\,
    \dot{p}=-\omega q-2\gamma\sqrt{1-j^{2}_{z}}\cos\phi,\\
   \dot{\phi}=\omega_{0}+\eta_{z}j_{z}-j_{z}(\eta_{x}\cos^{2}\phi+\eta_{y}\sin^{2}\phi)
   %\\ \nonumber
    -\frac{2\gamma j_{z}}{\sqrt{1-j^{2}_{z}}}q\cos\phi,\\
    \dot{j_{z}}=\left(1-j^{2}_{z}\right)(\eta_{x}-\eta_{y})\cos\phi\sin\phi+
    %\\ \nonumber
    2\gamma\sqrt{1-j^{2}_{z}}q\sin\phi.
\end{gather}

%%%%%%%%%%%%%%%%%%%%%%%%%%%%%%%%%%%%%%%%%%
%%%%%%%%%%%%%%%%%%%%%%%%%%%%%%%%%%%%%%%%%%
\section{Variables to plot energy surfaces}
\label{app:b}
%%%%%%%%%%%%%%%%%%%%%%%%%%%%%%%%%%%%%%%%%%
%%%%%%%%%%%%%%%%%%%%%%%%%%%%%%%%%%%%%%%%%%
We employ a new set of variables $u$ and $v$ associated with the qubit subspace to visualize better the energy surfaces: 
\begin{gather}
u=\arccos(-j_{z})\cos\phi,\,\,\,v=\arccos(-j_{z})\sin\phi,
\end{gather}
being the inverse transformation
\begin{gather}
j_{z}=-\cos\sqrt{u^{2}+v^{2}},\\ \nonumber
j_{x}=\frac{u}{\sqrt{u^{2}+v^{2}}}\sin\sqrt{u^{2}+v^{2}},\\ \nonumber
j_{y}=\frac{v}{\sqrt{u^{2}+v^{2}}}\sin\sqrt{u^{2}+v^{2}}.
\end{gather}
The $u$ and $v$ variables correspond to the angles $\phi$ and $\theta=\sqrt{u^{2}+v^{2}}$, i.e., the zenithal angle measured with respect to the pole. Then, we eliminate the bosonic variables $q$ and $p$ by employing Hamilton Eqs.~\ref{eq:he1} and~\ref{eq:he2}. Therefore, we obtain the energy surfaces ($\epsilon=E/\omega_{0}$) only as a function of the new variables $(u,v)$
\begin{gather}
\epsilon(\xi,u,v)=\sin^{2}\sqrt{u^{2}+v^{2}}\frac{1}{2\left(u^{2}+v^{2}\right)}\mbox{x}\\ \nonumber
\left[u^{2}\left(\frac{\eta_{x}}{\omega_{0}}-f_{\xi +}\right)\right.\left.+v^{2}\left(\frac{\eta_{y}}{\omega_{0}}-f_{\xi -}\right)\right]-\\ \nonumber \cos\sqrt{u^{2}+v^{2}}\left(1-\frac{\eta_{z}}{2\omega_{0}}\cos\sqrt{u^{2}+v^{2}}\right)
\end{gather}
with $f_{\xi\pm}=\gamma^{2}/\gamma_{\xi\pm}^{c}$. 

%%%%%%%%%%%%%%%%%%%%%%%%%%%%%%%%%%%%%%%%%%
%%%%%%%%%%%%%%%%%%%%%%%%%%%%%%%%%%%%%%%%%%
\section{Hessian matrix}
\label{app:c}
%%%%%%%%%%%%%%%%%%%%%%%%%%%%%%%%%%%%%%%%%%
%%%%%%%%%%%%%%%%%%%%%%%%%%%%%%%%%%%%%%%%%%

Here, we offer general expressions for the Hessian matrix of the system. Using the Hamilton from the main text we calculate that
\begin{gather}
 D^{(\xi)}(q,p,j_{z},\phi)=
 \left(
 \begin{array}{cccc}
 \frac{\partial^{2}H_{cl}^{(\xi)}}{\partial^{2}p} & \frac{\partial^{2}H_{cl}^{(\xi)}}{\partial q\partial p} & \frac{\partial^{2}H_{cl}^{(\xi)}}{\partial j_{z}\partial p} & \frac{\partial^{2}H_{cl}^{(\xi)}}{\partial \phi \partial p} \\
  \frac{\partial^{2}H_{cl}^{(\xi)}}{\partial p \partial q} & \frac{\partial^{2}H_{cl}^{(\xi)}}{\partial^{2} q} & \frac{\partial^{2}H_{cl}^{(\xi)}}{\partial j_{z}\partial q} & \frac{\partial^{2}H_{cl}^{(\xi)}}{\partial \phi \partial q} \\
  \frac{\partial^{2}H_{cl}^{(\xi)}}{\partial p \partial j_{z}} & \frac{\partial^{2}H_{cl}^{(\xi)}}{\partial q \partial j_{z}} & \frac{\partial^{2}H_{cl}^{(\xi)}}{\partial^{2} j_{z}} & \frac{\partial^{2}H_{cl}^{(\xi)}}{\partial \phi \partial j_{z}} \\
 \frac{\partial^{2}H_{cl}^{(\xi)}}{\partial p \partial \phi} & \frac{\partial^{2}H_{cl}^{(\xi)}}{\partial q \partial \phi} & \frac{\partial^{2}H_{cl}^{(\xi)}}{\partial j_{z}\partial \phi} & \frac{\partial^{2}H_{cl}^{(\xi)}}{\partial \phi^{2}}
 %%%
 \end{array}
 \right).
 \end{gather}
The second derivatives of the classical Hamiltonian are
\begin{gather}
\frac{\partial^{2} H_{cl}^{(\xi)}}{\partial p^{2}}=\frac{\partial^{2} H_{cl}^{(\xi)}}{\partial q^{2}}=\omega, \,\,\,\,
\frac{\partial^{2} H_{cl}^{(\xi)}}{\partial q\partial p}=\frac{\partial^{2} H_{cl}^{(\xi)}}{\partial p\partial q}=0,\\
\frac{\partial^{2} H_{cl}^{(\xi)}}{\partial j_{z}\partial p}=
\frac{\partial^{2} H_{cl}^{(\xi)}}{\partial p\partial j_{z}}=\frac{\gamma j_{z}}{\sqrt{1-j_{z}^{2}}}(1-\xi)\sin\phi\\
\frac{\partial^{2} H_{cl}^{(\xi)}}{\partial j_{z}\partial q}=\frac{\partial^{2} H_{cl}^{(\xi)}}{\partial q\partial j_{z}}=-\frac{\gamma j_{z}}{\sqrt{1-j_{z}^{2}}}(1+\xi)\cos\phi,\\
\frac{\partial^{2} H_{cl}^{(\xi)}}{\partial \phi\partial p}=\frac{\partial^{2} H_{cl}^{(\xi)}}{\partial p\partial \phi}=-\gamma\sqrt{1-j_{z}^{2}}(1-\xi)\cos\phi,\\
\frac{\partial^{2} H_{cl}^{(\xi)}}{\partial \phi\partial q}=\frac{\partial^{2} H_{cl}^{(\xi)}}{\partial q\partial \phi}=-\gamma\sqrt{1-j_{z}^{2}}(1+\xi)\sin\phi,\\
\frac{\partial^{2} H_{cl}^{(\xi)}}{\partial \phi\partial j_{z}}=\frac{\partial^{2} H_{cl}^{(\xi)}}{\partial j_{z}\partial \phi}=
j_{z}\left(\eta_{x}-\eta_{y}\right)2\cos\phi\sin\phi\\\nonumber+\frac{\gamma j_{z}}{\sqrt{1-j_{z}^{2}}}\left[(1+\xi)q\,\sin\phi+(1-\xi)p\,\cos\phi\right],\\
\frac{\partial^{2} H_{cl}^{(\xi)}}{\partial j_{z}^{2}}=\left[\eta_{z}-\left(\eta_{x}\cos^{2}\phi+\eta_{y}\sin^{2}\phi\right)\right]\\\nonumber-\frac{\gamma}{\left(1-j_{z}^{2}\right)^{3/2}}\left[(1+\xi)q\,\cos\phi-(1-\xi)p\,\sin\phi\right], \\
\frac{\partial^{2} H_{cl}^{(\xi)}}{\partial \phi^{2}}=\left(1-j_{z}^{2}\right)\left(\eta_{x}-\eta_{y}\right)\left(\sin^{2}\phi-\cos^{2}\phi\right)\\\nonumber-\gamma\sqrt{1-j_{z}^{2}}\left[(1+\xi)q\,\cos\phi-(1-\xi)p\,\sin\phi\right].
\end{gather}

The determinant of the Hessian matrix becomes
\begin{gather}
 D^{(\xi)}(q,p,j_{z},\phi)=\omega^{2}\left[\frac{\partial^{2}H_{cl}^{(\xi)}}{\partial j_{z}^{2}}\frac{\partial^{2}H_{cl}^{(\xi)}}{\partial \phi^{2}}-\left(\frac{\partial^{2}H_{cl}^{(\xi)}}{\partial j_{z}\partial\phi}\right)^{2}\right]\\\nonumber-\gamma^{2}j_{z}\left(1-\xi\right)\left(1+\xi\right)+
 \\ \nonumber
-\omega\gamma^{2}\left\{\left(1-j_{z}^{2}\right)\left[(1-\xi)^{2}\cos^{2}\phi+(1+\xi)^{2}\sin^{2}\phi\right]\frac{\partial^{2}H_{cl}^{(\xi)}}{\partial j_{z}^{2}}+\right.\\ \nonumber
\left.\frac{j_{z}^{2}}{1-j_{z}^{2}}\left[(1-\xi)^{2}\sin^{2}\phi+(1+\xi)^{2}\cos^{2}\phi\right]\frac{\partial^{2}H_{cl}^{(\xi)}}{\partial\phi^{2}}\right.+\\ \nonumber
\left.2 j_{z}\cos\phi\sin\phi\left[(1-\xi)^{2}-(1+\xi)^{2}\right]\frac{\partial^{2}H_{cl}^{(\xi)}}{\partial j_{z}\partial\phi}\right\}
\end{gather}

As an example, the Hessian at the points with $j_{zs}=\pm 1$, and $q_{s}=p_{s}=0$ is
\begin{gather*}
    D^{(\xi)}(0,0,\pm 1,\phi_{s})=-\omega^{2}(\eta_{x}-\eta_{y})^{2}\sin^{2}2\phi_{s}+\\ \nonumber
    \omega\gamma^{2}(\eta_{x}-\eta_{y})\left[\left((1+\xi)^{2}\cos^{2}\phi_{s}+(1-\xi)^{2}\sin^{2}\phi_{s}\right)\cos2\phi_{s}\right.\\\nonumber\left.-4\xi\sin^{2}2\phi_{s}\right] 
    +\gamma^{4}(1-\xi)^{2}(1+\xi)^{2},
\end{gather*}
If we consider the symmetric case $\xi=0$ when $\eta_{x}=\eta_{y}$, where the rotational symmetry is obtained one gets $D^{0}_{\eta_{x}=\eta_{y}}(0,0,\pm 1,\phi_{s})=\gamma^{4}$. Thus, the points at $j_{zs}=1\pm$ must be either a maximum or a minimum. A simple inspection of the energy surfaces reveals their nature, which is confirmed if one calculates the spectrum of the Hessian matrix. 
    
%%%%%%%%%%%%%%%%%%%%%%%%%%%%%%%%%%%%%%%%%%
\subsection{Hessian determinant in the Tavis-Cummings limit}
%%%%%%%%%%%%%%%%%%%%%%%%%%%%%%%%%%%%%%%%%%

For the Tavis-Cummings limit ($\xi=0$), the determinant of the Hessian takes the form
\begin{gather}
 D^{(0)}(q,p,j_{z},\phi)=\omega^{2}\left[\frac{\partial^{2}H_{cl}^{(0)}}{\partial j_{z}^{2}}\frac{\partial^{2}H_{cl}^{(0)}}{\partial \phi^{2}}-\left(\frac{\partial^{2}H_{cl}^{(0)}}{\partial j_{z}\partial\phi}\right)^{2}\right]\\\nonumber-\gamma^{2}j_{z}
-\omega\gamma^{2}\left[\left(1-j_{z}^{2}\right)\frac{\partial^{2}H_{cl}^{(0)}}{\partial j_{z}^{2}}+\frac{j_{z}^{2}}{1-j_{z}^{2}}\frac{\partial^{2}H_{cl}^{(0)}}{\partial\phi^{2}}\right].
\end{gather}
with
\begin{gather}
\frac{\partial^{2} H_{cl}^{(0)}}{\partial j_{z}\partial \phi}=
j_{z}\left(\eta_{x}-\eta_{y}\right)2\cos\phi\sin\phi\\\nonumber+\frac{\gamma j_{z}}{\sqrt{1-j_{z}^{2}}}\left(q\,\sin\phi+p\,\cos\phi\right),\\
\frac{\partial^{2} H_{cl}^{(\xi)}}{\partial j_{z}^{2}}=\left[\eta_{z}-\left(\eta_{x}\cos^{2}\phi+\eta_{y}\sin^{2}\phi\right)\right]\\\nonumber-\frac{\gamma}{\left(1-j_{z}^{2}\right)^{3/2}}\left(q\,\cos\phi-p\,\sin\phi\right), \\
\frac{\partial^{2} H_{cl}^{(0)}}{\partial \phi^{2}}=\left(1-j_{z}^{2}\right)\left(\eta_{x}-\eta_{y}\right)\left(\sin^{2}\phi-\cos^{2}\phi\right)\\\nonumber-\gamma\sqrt{1-j_{z}^{2}}\left(q\,\cos\phi-p\,\sin\phi\right).
\end{gather}

%%%%%%%%%%%%%%%%%%%%%%%%%%%%%%%%%%%%%%%%%%
\subsection{Hessian determinant in the Dicke limit}
%%%%%%%%%%%%%%%%%%%%%%%%%%%%%%%%%%%%%%%%%%

Instead, for the Dicke limit ($\xi=0$), the determinant of the Hessian becomes
\begin{gather}
 D^{(1)}(q,p,j_{z},\phi)=\omega^{2}\left[\frac{\partial^{2}H_{cl}^{(1)}}{\partial j_{z}^{2}}\frac{\partial^{2}H_{cl}^{(1)}}{\partial \phi^{2}}-\left(\frac{\partial^{2}H_{cl}^{(1)}}{\partial j_{z}\partial\phi}\right)^{2}\right]+ \\ \nonumber
-\omega\gamma^{2}\left\{\left(1-j_{z}^{2}\right)4\sin^{2}\phi\frac{\partial^{2}H_{cl}^{(1)}}{\partial j_{z}^{2}}\right.\\\nonumber+
\left.\frac{j_{z}^{2}}{1-j_{z}^{2}}4\cos^{2}\phi\frac{\partial^{2}H_{cl}^{(1)}}{\partial\phi^{2}}-8j_{z}\cos\phi\sin\phi\frac{\partial^{2}H_{cl}^{(1)}}{\partial j_{z}\partial\phi}\right\}.
\end{gather}
with
\begin{gather}
\frac{\partial^{2} H_{cl}^{(1)}}{\partial j_{z}\partial \phi}=
j_{z}\left(\eta_{x}-\eta_{y}\right)2\cos\phi\sin\phi+\\\nonumber\frac{2\gamma j_{z}}{\sqrt{1-j_{z}^{2}}}q\,\sin\phi,\\
\frac{\partial^{2} H_{cl}^{(1)}}{\partial j_{z}^{2}}=\left[\eta_{z}-\left(\eta_{x}\cos^{2}\phi+\eta_{y}\sin^{2}\phi\right)\right]\\\nonumber-\frac{2\gamma q\,\cos\phi}{\left(1-j_{z}^{2}\right)^{3/2}}, \\
\frac{\partial^{2} H_{cl}^{(1)}}{\partial \phi^{2}}=\left(1-j_{z}^{2}\right)\left(\eta_{x}-\eta_{y}\right)\left(\sin^{2}\phi-\cos^{2}\phi\right)\\\nonumber-2\gamma\sqrt{1-j_{z}^{2}}q\,\cos\phi.
\end{gather}

%%%%%%%%%%%%%%%%%%%%%%%%%%%%%%%%%%%%%%%%%%
%\reftitle{References}
%\externalbibliography{yes}
\bibliography{Dicke22_v2}
%%%%%%%%%%%%%%%%%%%%%%%%%%%%%%%%%%%%%%%%%%

%%%%%%%%%%%%%%%%%%%%%%%%%%%%%%%%%%%%%%%%%%
\end{document}